\documentclass[prb,twocolumn,letterpaper, superscriptaddress,longbibliography]{revtex4-2}
\usepackage{graphicx}
\usepackage{dcolumn}
\usepackage{bm}
\usepackage{color}
\usepackage{tabularx}
\usepackage{array}

\usepackage[utf8]{inputenc}
\usepackage[T1]{fontenc}
\usepackage{color,soul}
\usepackage{tabularx}
\usepackage{amsmath}

\begin{document}

\preprint{APS/123-QED}

\title{Perspective: Magnon-magnon coupling in hybrid magnonics}

\author{Wei Zhang}
\email{All correspondance should be addressed to: zhwei@unc.edu}
\affiliation{Department of Physics and Astronomy, University of North Carolina at Chapel Hill, Chapel Hill, NC 27599, USA}

\author{Yuzan Xiong}
\affiliation{Department of Physics and Astronomy, University of North Carolina at Chapel Hill, Chapel Hill, NC 27599, USA}

\author{Jia-Mian Hu}
\affiliation{Department of Materials Science and Engineering, University of Wisconsin-Madison, Madison, Wisconsin, 53706, USA}

\author{Joseph Sklenar}
\affiliation{Department of Physics and Astronomy, Wayne State University, Detroit, MI 48202, USA}

\author{Mitra Mani Subedi}
\affiliation{Department of Physics and Astronomy, Wayne State University, Detroit, MI 48202, USA}

\author{M.Benjamin Jungfleisch}
\affiliation{Department of Physics and Astronomy, University of Delaware, Newark, DE 19716, USA}

\author{Vinayak S. Bhat}
\affiliation{Department of Physics and Astronomy, University of Delaware, Newark, DE 19716, USA}

\author{Yi Li}
\affiliation{Materials Science Division, Argonne National Laboratory, Argonne, IL 60439, USA}

\author{Luqiao Liu}
\affiliation{Department of Electrical Engineering and Computer Science, Massachusetts Institute of Technology, Cambridge, Massachusetts 02139, USA}

\author{Qiuyuan Wang}
\affiliation{Department of Electrical Engineering and Computer Science, Massachusetts Institute of Technology, Cambridge, Massachusetts 02139, USA}

\author{Yunqiu Kelly Luo}
\affiliation{Department of Physics and Astronomy, Department of Chemistry, Mork Family Department of Chemical Engineering and Materials Science, University of Southern California, Los Angeles, CA, 90089 USA}

\author{Youn Jue Bae}
\affiliation{Department of Chemistry and Chemical Biology, Cornell University, Ithaca, NY, 14850, USA}

\author{Benedetta Flebus}
\affiliation{Department of Physics, Boston College, 140 Commonwealth Avenue Chestnut Hill, Massachusetts 02467, USA}

\begin{abstract}  
\centering{\textbf{ABSTRACT}} \\

The internal coupling of magnetic excitations (magnons) with themselves has created a new research sub-field in hybrid magnonics, i.e., magnon-magnon coupling, which focuses on materials discovery and engineering for probing and controlling magnons in a coherent manner. This is enabled by, one, the abundant mechanisms of introducing magnetic interactions, with examples of exchange coupling, dipolar coupling, RKKY coupling, and DMI coupling, and two, the vast knowledge of how to control magnon band structure, including field and wavelength dependences of frequencies, for determining the degeneracy of magnon modes with different symmetries. In particular, we discuss how magnon-magnon coupling is implemented in various materials systems, with examples of magnetic bilayers, synthetic antiferromagnets, nanomagnetic arrays, layered van der Waals magnets, and (DMI SOT materials) in magnetic multilayers. We then introduce new concept of applications for these hybrid magnonic materials systems, with examples of frequency up/down conversion and magnon-exciton coupling, and discuss what properties are desired for achieving those applications.
\end{abstract}

\flushbottom
\maketitle

\section*{Introduction}

Hybrid solid-state systems play an important role in experimental and theoretical condensed matter physics. They harness interacting excitations, such as sound waves (phonons), microwave (MW) and light waves (photons), and quantum defects (spin color centers) to complete tasks that are beyond the capability of each individual system \cite{kurizki2015quantum,lachance2019hybrid}. Recent development in quantum science has put such hybrid systems to the forefront of condensed matter physics research, calling for upgraded benchmark performance in the efficient, coherent, and robust transformation of information carried by these fundamental excitations.

\begin{figure*}[htb]
\centering
\includegraphics[width=6.7 in]{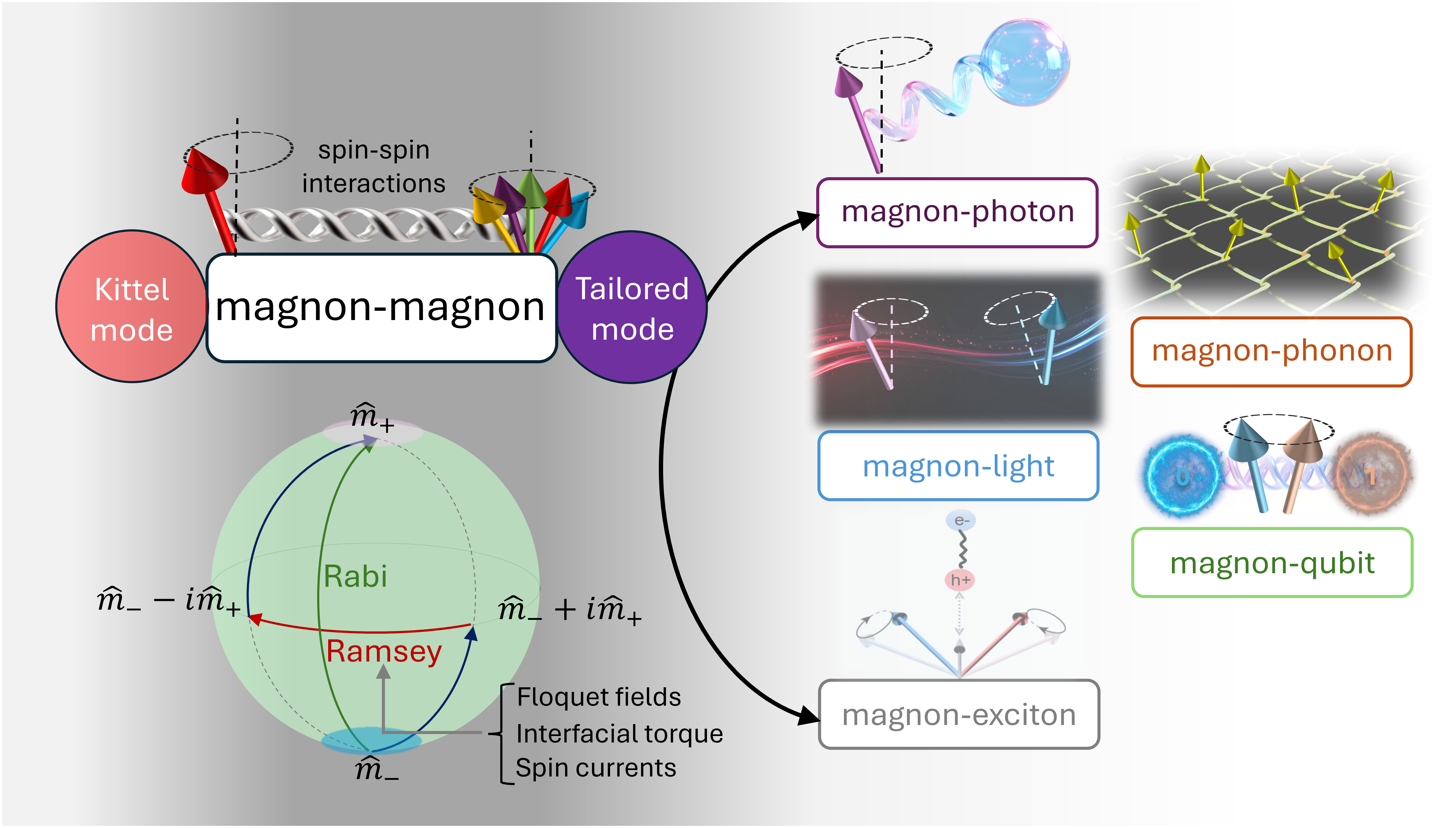}
\caption{The field of hybrid magnonics has grown and developed into different ramifications, such as magnon-photon, magnon-phonon, and magnon-light coupling systems, with the goal of developing energy and signal transduction functionalities across different physical platforms. The magnon-magnon coupling is presented as a unique and versatile approach towards inducing and tailoring magnon modes with desirable properties for further hybridization in the various hybrid magnonic contexts, i.e. Magnon + $X$. \textcolor{black}{The Bloch sphere representation shows how hybrid magnonics can perform coherent information processing, e.g. Rabi oscillation and Ramsey interference, as analogous to a two-level system \cite{zhang2019electronically,zhuang2024dynamical}. The symbols $\hat{m}_+$ and $\hat{m}_-$ represent some generic single magnon modes (states).} }
\label{fig_hm}
\end{figure*} 
 
Spin waves (or magnons), have received increased attention in such hybrid quantum systems due to their small wavelength, dissipationless transport, nonlinearity, GHz-THz resonance frequency, and notably, the ability to hybridize with a wide variety of other quasiparticles and light waves, hence the concept of ``hybrid magnonics'' \cite{AwschalomPRXQ21,LiJAPReview2020,yuan2022quantum,flebus20242024,chumak2022advances}. To date, strong and coherent couplings between magnons and other excitations, such as microwave photons, light photons, phonons, single spins, and qubits, have been demonstrated. Consequently, the field of hybrid magnonics has grown and developed into different ramifications, such as magnon-photon, magnon-phonon, and magnon-light coupling systems, etc, presenting a broad spectrum of coherent phenomena, including but not limited to, level repulsion/attraction \cite{goryachev2014high,zhang2014strongly,tabuchi2014hybridizing,wang2020dissipative,harder2018level}, magnetically-induced transparency(MIT) \cite{xiong2020probing,xiong2022tunable,xiong2024phase}, super/ultrastrong couplings \cite{inman2022hybrid}, pump-induced nonlinearity \cite{rao2023unveiling,yang2025control,xiong2024magnon,qu2025pump,wu2025coupling}, zero-reflection(ZR) \cite{qian2023non,christy2025tuning}, and spectrum singularities \cite{yang2020unconventional,zhang2019experimental,xiong2025photon}, bestowing emerging quantum engineering functionalities \cite{lachance2019hybrid,LiJAPReview2020,yuan2022quantum,flebus20242024,chumak2022advances}.

In hybrid magnonics, so far, most attention has been focused in hybridizing magnons with another excitation of distinct nature, i.e., Magnon + $X$, with the goal of developing energy and signal transduction functionalities across different physical platforms \textcolor{black}{and coherent information processing in an artificial two-level system. For example, treating the hybrid system as an analogous ``two-level'' system in the sense of the Jaynes–Cummings Hamiltonian, the Rabi process shuffles the two-level states in an analogous Bloch sphere representation, shown in Fig. \ref{fig_hm}. On the magnon side, the Kittel (wavevector ${\bf k}=0$) mode, being the most prominent and zero-th order, was often exploited, see Fig. \ref{fig_hm}. However, introducing finite-${\bf k}$ modes may bring in further advantages, such as using the wavevector ${\bf k}$ as a state variable, leveraging the full magnon dispersion, and integrating the magnon precessional phase ($\phi$) as a control parameter into the technologically important Ramsey process.} \textcolor{black}{Computational demonstration of such a ``magnonic Ramsey process'' has recently been reported in magnon-photon systems \cite{zhuang2024dynamical} using protocols that are similar to classical-domain demonstration of the Ramesy process in a two-level photonic system \cite{zhang2019electronically}. Such a scheme can be also extended to magnon-magnon coupled systems \textcolor{black}{between two generic single magnon modes ($\hat{m}_+$ and $\hat{m}_-$)}, in which additional, novel control schemes can be introduced, including interface exchange, spin current (spin torque), and Floquet field, due to both subsystems being “magnonic” in nature.}

On the other hand, underpinning the Magnon + $X$ hybridization is the fundamental versatile ``spin-couplings'' manifesting magneto-optical, magneto-electric, and magneto-strictive effects \cite{lachance2019hybrid}. Besides the cross-platform Magnon + $X$ hybridizations, one distinct advantage of magnons is their strong/coherent interaction among themselves -- the magnon-magnon coupling -- providing a unique route towards generation of novel, tailored magnon modes \cite{fan2020manipulation,klingler2018spin,chen2018strong,qin2018exchange,li2020coherent,xiong2020probing,li2024reconfigurable,liu2024strong,lukas2019ferri,macneill2019gigahertz,cham2022anisotropic,diederich2023tunable,li2023ultrastrong,makihara2021ultrastrong,zhang2024terahertzupconversion,zhang2024terahertznonlinear}. Such magnon-magnon interaction underpins a diverse range of coherent phenomena that are governed by spin-spin interactions, including but not limited to, direct exchange \cite{coey2010magnetism}, Ruderman–Kittel–Kasuya–Yosida (RKKY) \cite{ruderman1954indirect}, and Dzyaloshinskii–Moriya interaction (DMI) \cite{dzyaloshinsky1958thermodynamic,moriya1960anisotropic}. The different symmetries in these couplings can be synergistically combined to generate, modulate, and detect unconventional mode forms in a Magnon + $X$ hybrid, thereby providing tailored magnon modes as further candidates for multi-partite, cross-platform hybridizations as illustrated in Fig. 1. Specifically, integrating magnon-magnon coupling in the Magnon + $X$ process offers several derived advantages, as summarized in Fig.\ref{fig_attributes}, Keys 1 -- 6:

\begin{figure}[htb]
\centering
\includegraphics[width=3.4 in]{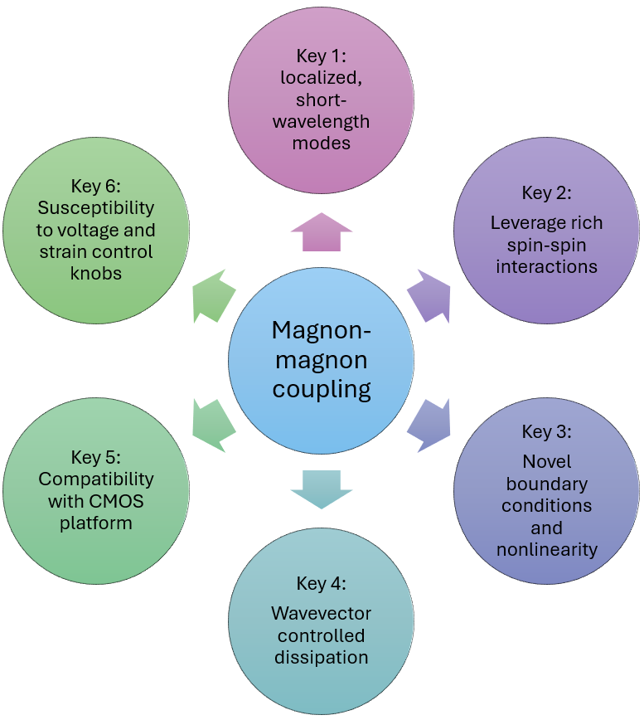}
\caption{The various key attributes of magnon-magnon coupling in the context of hybrid magnonics. }
\label{fig_attributes}
\end{figure}

\noindent 1. \textit{Wavelength-selectivity}: the interfacial spin couplings can be highly localized, hence allowing effective excitation of short-wavelength magnon modes down to the deep-exchange magnon regime \cite{xiong2020probing,xiong2022tunable,inman2022hybrid,chen2019excitation,chen2018strong,klingler2018spin,li2020coherent}. \textcolor{black}{Such an interfacial spin perturbation scheme is analogous to the excitation of coherent THz magnons in ferromagnet thin films by current-induced interfacial spin torques \cite{razdolski2017nanoscale,BrandtPhysRevB2021,salikhov2023coupling}.} 
 
\noindent 2. \textit{Superimposing spin-spin interactions}: the various spin-coupling types, direct-Ex, RKKY, dipolar, and DMI, entail different symmetries and controlling parameters. Therefore, they can be synergistically engineered to interfere, constructively or destructively for selective mode excitations \cite{subedi2024even,subedi2025investigating,comstock2023hybrid,liang2023ruderman,wang2024ultrastrong,wang2024ultrastrong1}. \textcolor{black}{For example, in a canonical magnon-magnon coupled Permalloy/YIG bilayer structure, it has recently been shown \cite{christy2025tuning} that the phase of magnetization precession at the interface is collectively controlled by both the Zeeman torque (from the microwave drive) and the interlayer exchange torque. Such a dual-channel excitation scheme provides a new knob to control both the phase and magnetization amplitude of the individual exchange magnon modes in the YIG film, suggesting new opportunities in the realm of phase-controlled hybrid magnonoics.}

\noindent 3. \textcolor{black}{\textit{Novel boundary conditions and strong nonlinearity}: beyond the extended, pristine films, new boundary conditions to the excited magnon modes can be imposed by nano-structuring (such as in the cases of artificial spin-ice and magnonic crystals) \cite{Sultana_2025,chumak2017magnonic}, interfacing with other layers \cite{hellman2017interface}, additional current injections \cite{manchon2019current}, and mutual spin pumping \cite{li2020coherent,subedi2025engineering}, hence allowing to leverage the rich collective magnonic dispersions and their high tunability  \cite{an2024emergent,pan2024magnon,zheng2023tutorial,schonfeld2025dynamical,xiong2024magnon,li2023floquet}. Besides, the strong nonlinearity of magnonics can be used to enrich hybrid magnon modes, leading to finite mode-overlap of different magnon modes that are presumed to be orthogonal in the linear regime. Examples include the recently discovered pump-induced mode anticrossing arising from the three-magnon scattering \cite{rao2023unveiling,qu2025pump}, opto-electro-magnonic oscillator due to nonlinear magnon-photon coupling \cite{xiong2024magnon}, and coherent magnon mode generation in the nonlinear excitation regime \cite{an2024emergent,pan2024magnon,zheng2023tutorial,schonfeld2025dynamical,li2023floquet}.}


\noindent 4. \textit{Dissipation(damping) control}: \textcolor{black}{magnon-magnon coupling can modify effective magnetic damping of a mode through mutual spin pumping, especially when two modes are at same or nearly same frequencies (i.e. at co-resonance or near co-resonance condition), which offers two major benefits: (i) reducing energy loss, which can increase the distance waves can travel and hence improve energy efficiency, and (ii) strengthening or weakening (selectively) specific modes, which can be used to stabilize useful oscillations while filtering unwanted ones. These capabilities would be highly valuable for spintronic and magnonic device applications.  For example, by controlling the damping of individual modes electrical control of magnon frequencies in the vicinity of an avoided energy level crossing is enabled \cite{li2020coherent,subedi2025engineering,fan2025dynamically,schmoll2025wavenumber,serha2024magnetic,wang2024broad,sklenar2021self}.}

\noindent 5. \textit{Engineering scalability}: magnon coupling involves, more often than not, magnetic films and multilayers on planar substrates (including wafers), a modular component that shares the same physical form factor and packing model as those of silicon, thus is reconcilable with the needs for scaling with state-of-the-art CMOS architectures, \textcolor{black}{as opposed to conventional resonator platforms like YIG spheres and 3D cavities}. \textcolor{black}{Recent advances in magnonics also advocate a scheme shift from large magnetic volumes to nanoscale lithographic structures, to realize miniaturization and true CMOS compatibility rather than relying on stand-alone bulk resonators. Concrete progress has already aligned with the CMOS style footprints. For instance, fully planar lithographic demonstrations show reconfigurable magnon-magnon coupling in ultrathin insulators under a spin-orbit torque control: Pt nanostripes patterned atop 3-nm Bi-doped YIG define on-chip magnonic cavities whose modes hybridize with boundary magnons, yielding a tunable anticrossing gap and enabling electrical on/off control of the coupling \cite{wang2025current}. The compact, planar geometry supports multiple cavities and engineerable inter-cavity couplings, the kind of footprint that naturally maps to CMOS packaging. As efficient magnon-based computing blocks have been experimentally demonstrated, with the additional versatility provided by magnon-magnon coupling, a hybrid computing architecture with additional functionality and/or efficiency can be anticipated \cite{wang2025current,barman2025external,wang2024nanoscale}.}

\noindent 6. \textit{Compatible with mature spintronic/magnetic control knobs}: since hybrid magnon-magnon coupling can be achieved with interfacial coupling in magnetic thin-film bilayers \cite{fan2025dynamically,li2024reconfigurable,li2020coherent,wang2025current,sud2025electrically,li2023floquet,freeman2025tunable}, \textcolor{black}{conventional mature spintronic and magnetic control toolkits, such as spin-transfer torque \cite{ChenIEEE10}, spin-orbit torque \cite{ShaoIEEE21}, strain control \cite{SadovnikovPRL18} and voltage control \cite{DaiJMMM22}, can be directly adopted for engineering magnon-magnon couplings}. \textcolor{black}{However, the main engineering bottleneck remains the magnon electron transducer. The efficiency of the usual conversion routes based on microwave antennas for excitation and spin pumping with inverse spin Hall readout is low. This limits the density of addressable circuits at the chip level \cite{barman2025external}. A promising route is the magnetoelectric transducer that combines piezoelectric and magnetostrictive elements. This approach is driven by ac voltage and is expected to be far more energy efficient than any current-driven antennas. In optimized cells the projected operation can reach the atto-joule energy scale \cite{wang2025current}. } 

It is natural to expect that a growing ability to tune magnon–magnon interactions will reorient the role of spintronic systems -- adding new research thrusts atop predominantly application-driven goals toward their use as platforms for exploring emergent phenomena of fundamental interest. For instance, the programmable control over dispersion and hybridization of synthetic antiferromagnets (syn-AFMs)~\cite{Duine2018syntheticAFM}, whose collective spin-wave modes can be systematically reconfigured through interlayer exchange, anisotropy, and symmetry-breaking fields,  draws a natural parallel to how metamaterials have been used to access novel regimes of dynamics and topology~\cite{lu2023non,ozawa2019topological}. As both the coherent and incoherent of their long-wavelength behavior is well understood, magnon-magnon coupled syn-AFMs might offer accessible framework for investigating the emergence of non-Hermitian phenomena that have been attracting so much attention in recent years~\cite{hurst2022non}.

From a fundamental point of view, magnonic systems provide a rare—though largely unexplored—opportunity to directly test the resilience of non-Hermitian dynamics in the presence of strong intrinsic nonlinearities~\cite{gunnink2022nonlinear,deng2023exceptional}, an aspect largely inaccessible in other platforms. At the same time, the application of these principles may also unlock new functionalities, such as reconfigurable spintronic diodes based on nonreciprocal spin-wave transport—particularly as progress continues in engineering chiral interactions within magnetic heterostructures~\cite{jamali2013spin}. 

Another intriguing, yet still underexplored,  in magnonic systems is Floquet engineering~\cite{Oka2018FloquetReview}, where periodic driving can induce new modes, topological band structures, or symmetry-selective coupling. While synthetic antiferromagnets offer a natural setting for this approach~\cite{li2023floquet}, equally promising are the discrete spin-wave modes that arise in laterally confined magnetic structures, which can serve as building blocks for implementing synthetic dimensions or dynamically tunable coupling networks~\cite{yu2025comprehensive,rudner2020band}. 
Together, these examples highlight just a few of the opportunities that become accessible as magnon–magnon interactions are brought under experimental control, suggesting that a broad and largely untapped landscape of driven, correlated, and topologically structured magnonic phenomena lies within reach.

\begin{figure*}[htb]
\centering
\includegraphics[width=6.7 in]{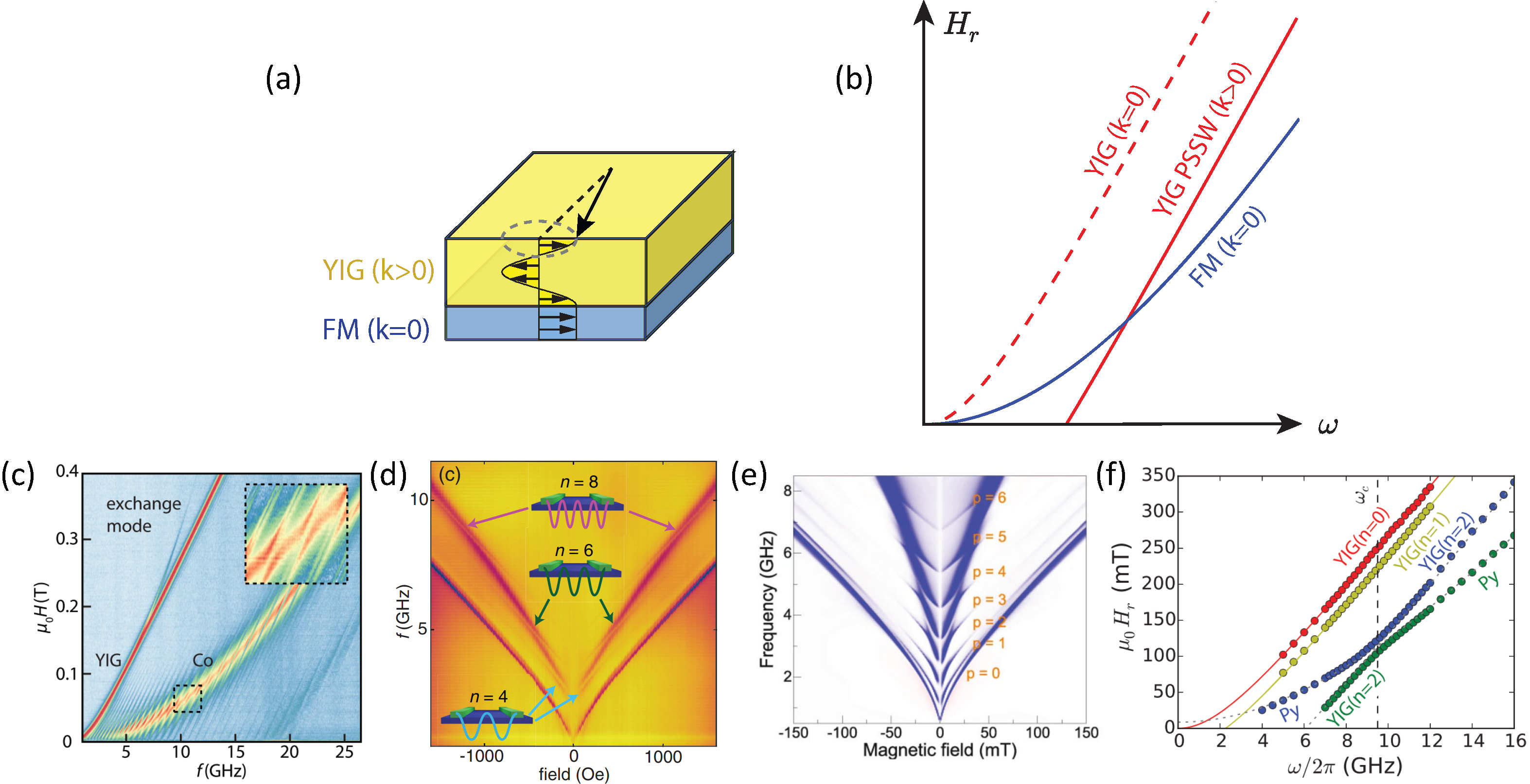} 
\caption{(a) Schematic of coupled magnon-magnon dynamics in a YIG/FM bilayer mediated by interfacial exchange coupling, with uniform mode ($k=0$) excited in the FM layer and the perpendicular standing spin wave mode ($k>0$) excited in the YIG layer. (b) $H_r$-$\omega$ dependence for the YIG PSSW mode (red) and the FM uniform mode (blue) that intersect with each other. YIG uniform mode is also shown in red dashed curve. (c-f) Coherent magnon-magnon coupling in (c) YIG/Co \cite{klingler2018spin}, (d) YIG/Ni\cite{chen2018strong}, (e) YIG/CoFeB\cite{qin2018exchange} and (f) YIG/NiFe\cite{li2020coherent}.}
\label{fig_yig1}
\end{figure*}

\section*{Magnon-magnon coupling enabled by different types of spin-spin interactions}

Below, we review the subject based on the different types of spin-spin interactions pertinent to each magnon-magnon system category, a comprehensive list of material examples can be found in Table \ref{Tab:list}.   

\subsection*{Interfacial exchange coupling in magnetic bilayers} 

The study of coherent dynamic coupling between two adjacent magnetic layers was among the earliest explorations of coherent magnon-magnon coupling, motivated primarily by the question of how to efficiently excite short-wavelength exchange spin waves. Usually, one needs to create a spatially nonuniform microwave field to achieve geometrical overlap with the spin wave, enabling their coupling and energy exchange. This typically was done by fabricating nanoscale microwave antennas, which, however, further reduces coupling efficiency due to the narrow antenna electrodes.

\paragraph{Metal-insulator system}

In 2018, Klinger \textit{et al}. \cite{klingler2018spin}, Chen \textit{et al}. \cite{chen2018strong}, and Qin \textit{et al}. \cite{qin2018exchange} reported the excitation of the perpendicular standing spin wave (PSSW) mode--an exchange spin wave mode with a wave vector perpendicular to the thin film plane--in a YIG film, assisted by an adjacent ferromagnetic (FM) layer, see \textcolor{black}{Fig.\ref{fig_yig1} (c-e)}. By growing the FM layer on top of YIG film, interfacial exchange coupling (dipolar coupling in Chen \textit{et al}.’s work) enables coherent magnetic excitation transfer from the FM layer to YIG \textcolor{black}{[Fig.\ref{fig_yig1} (a)]}. Since the selected FM materials (Co, Ni, or CoFeB) have much larger magnetization compared to YIG, the in-plane field dependence of their Kittel modes exhibits different $\omega-H$ slopes, allowing one mode to intersect with the other due to the PSSW frequency offset in YIG \textcolor{black}{[Fig.\ref{fig_yig1}(b), (f)]}. Moreover, because the Kittel mode in the FM layer efficiently couples with a uniform microwave rf field, the spatially nonuniform PSSW mode in YIG can be excited by the uniform microwave field through its interfacial interaction with the FM layer, making it easily observable experimentally. Without this technique, the PSSW mode barely couples to the uniform microwave field, with the amplitude of the $n=1$ PSSW mode being 50 times weaker than that of the uniform mode \cite{LiPRL16}.

\begin{figure*}[htb]
\centering
\includegraphics[width=6.5 in]{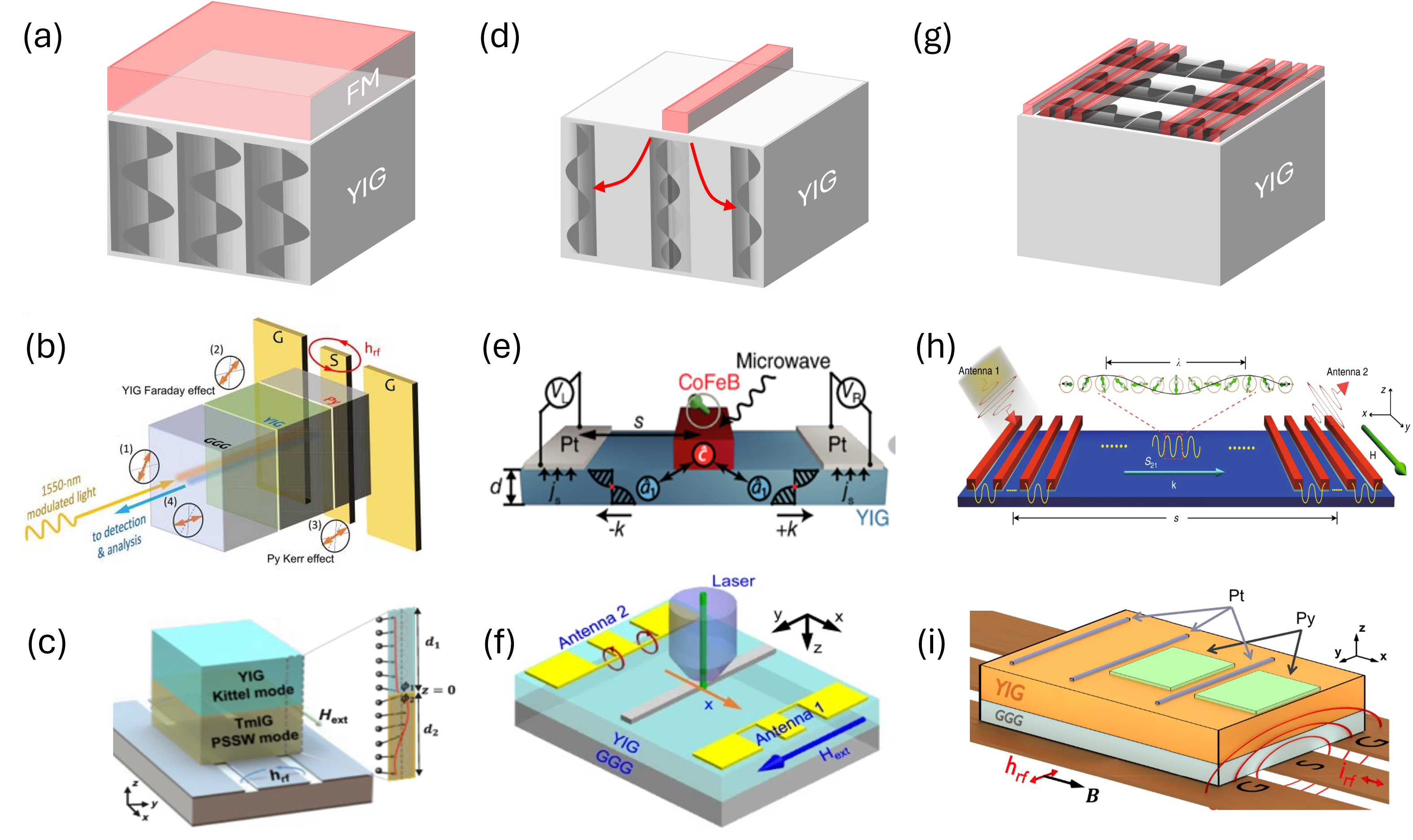}
\caption{The various geometries of magnon-magnon coupling in bilayer magnetic heterostructures. (a,b,c): uniform, extended bilayer coupling in which the thickness of the film defines the magnonic cavity, for (b) insulator/metal (Figure taken from Ref.\cite{xiong2020probing}), and (c) all-insulator systems (Figure taken from Ref. \cite{liu2024strong}). (d,e,f): the device geometry in which an FM stripe couples locally to an extended YIG film. The excited magnons can propagate along the lateral dimension, and be detected via: (e) dc rectifications, e.g., using spin Hall effect of Pt (Figure taken from Ref. \cite{sheng2023nonlocal}), or (f) with rf inductive antennas and magneto-optical probes (Figure taken from Ref. \cite{qin2021nanoscale}). (g,h,i): lateral magnonic cavity in which the FM stripes (h) serve as complementary, local antennas (Figure taken from Ref. \cite{liu2018long}), or (i) only used to define the cavity boundaries while the excitation is globally sourced otherwise (Figure taken from Ref. \cite{santos2023magnon}). }
\label{fig_yig2}
\end{figure*} 

Another significant consequence of these experiments is the observation of magnon mode hybridization between the FM and YIG layers, which has a profound impact \textcolor{black}{in their application in \textcolor{black}{coherent} information processing \cite{lachance2019hybrid,LiJAPReview2020}. This demonstrates that magnon excitations can be coherently exchanged between the two spatially separated magnetic systems, enabling a highly compact, solid-state hybrid magnonic system with a nanometer-scale total vertical stacking, in complementary to their lateral dimensions}. Since the coupling originates from the interfacial interaction between the two magnetic layers, the magnetic bilayer hybrid magnonic system allows for \textcolor{black}{structural engineering} along the lateral dimensions without affecting the properties of mode hybridization \textcolor{black}{\cite{LiAPL21,WaglePRB25}}, allowing room for integrating with additional designs and constructions at the chip level. 

Figure \ref{fig_yig2} summarizes the various geometries for magnon-magnon coupling investigations, pertinent to different types of magnon modes of interest. The left column [Fig.\ref{fig_yig2}(a,b,c)] shows uniform, extended bilayer coupling in which the thickness of the film defines the magnonic cavity. The situation applies to both insulator/metal [Fig.\ref{fig_yig2}(b)] and all-insulator [Fig.\ref{fig_yig2}(c)] systems. The middle column [Fig.\ref{fig_yig2}(d,e,f)] exemplifies the device geometry in which an FM stripe (via lithography, etc) couples locally to an extended YIG film. In this case, the magnon-magnon coupling leverages both the direct exchange coupling (at the interface) and an enhanced dipolar interaction (due to micro- or nano-structuring). The locally excited magnon modes (at and below the center stripe), instead of being confined by a cavity, can propagate along the lateral sides, and be detected via dc rectifications, e.g., using spin Hall effect of Pt, as in Fig.\ref{fig_yig2}(e), rf inductive antennas, or magneto-optical probes, see Fig.\ref{fig_yig2}(f). Further, the right column [Fig.\ref{fig_yig2}(g,h,i)] exemplifies the construction of a lateral magnonic cavity by using pairs of FM stripe(s). In such a case, the magnon modes are confined not only vertically but also laterally defined by the dimensions of the cavity, which can be easily controlled by the location of the stripe(s). Notably, the FM stripes can also serve as complementary, local antennas, as in Fig.\ref{fig_yig2}(h), apart from being the cavity boundaries; or else, they can be used only to define the cavity boundaries while the excitation is globally sourced by an external rf structure (e.g. a waveguide beneath).

\paragraph{Coupling strengths}

The interfacial exchange coupling provides an interlayer magnon-magnon coupling strength ($g_c$) which can be derived analytically from the Landau-Lifshitz-Gilbert equation, as derived by Li \textit{et al}.\cite{li2020coherent}:
\begin{equation}
 g_c \simeq \sqrt{{J\over M_1t_1}{J\over M_2t_2}}
\label{eq:J} 
\end{equation}
\noindent Here, $J$ is the interfacial coupling energy density between the two magnetic layers, while $M_{1,2}$, $t_{1,2}$ represent the magnetization and the thickness of the two layers, respectively. Note that there is a pre-factor in Eq.\ref{eq:J} which comes from the ellipticity of magnetization precession and is close to one \cite{li2020coherent}. Eq. \ref{eq:J} shows that the interlayer magnon-magnon coupling comes from a mutual interaction of the two layers, and their coupling strength is the geometric average of the effective interfacial exchange field, which is inversely proportional to the magnetization and thin film thickness. This is similar to the coupling strength in magnon-photon coupling, where the coupling strength is inversely proportional to the square root of the effective cavity volume and total magnetic moment (since the counterpart of $J$ is proportional to the total magnetic moment, the resulting magnon-photon coupling strength becomes proportional to the square root of the total magnetic moment). Note that $g_c$ gives the \textcolor{black}{resonance} field splitting. The frequency splitting, which represents the real magnon-magnon coupling strength, requires field-to-frequency conversion along the magnon dispersion curve. Because of the effective demagnetization field in the Kittel equation, the conversion factor is usually larger than the single-electron gyromagnetic ratio of 2.8 MHz/Oe.

The interfacial exchange coupling strength is determined by the electronic and lattice properties of the materials. Reported coupling strengths are $J=0.4$ mJ/m$^2$ for YIG/Co \cite{klingler2018spin}, and $0.06$ mJ/m$^2$ for YIG/Py \cite{li2020coherent}. \textcolor{black}{It is worth noting that as a back-to-back published paper with Ref. \cite{klingler2018spin}, a different mechanism, i.e. the dipole-dipole coupling of spin waves in YIG/Ni nanograting system, can also yield a similar interfacial magnon-magnon coupling strength of 0.03 mJ/m$^2$ \cite{chen2018strong}}. The YIG/CoFeB interface exhibits a magnon-magnon anticrossing gap \cite{qin2018exchange} similar to YIG/Py, suggesting that their $J$ values are of the same order of magnitude. It is noted that the interfacial exchange coupling strength is significantly smaller than the intrinsic exchange coupling in ferromagnets. For example, in permalloy (Ni$_{80}$Fe$_{20}$, or Py) the interfacial exchange energy can be expressed as $2A_{ex}/a$ where $A_{ex}$ is the exchange constant and $a$ is the lattice parameter \cite{HillebrandsPRB90}. Taking $A_{ex}=12$ pJ/m and $a=0.36$ nm for Py \cite{LiPRL16}, one finds $2A_{ex}/a=68$ mJ/m$^2$ which is 3 orders of magnitude larger than $J$. 

The interfacial exchange coupling between YIG and metallic FM is found to be antiferromagnetic (i.e., $J$ is negative) in most bilayer systems, using magnetometry and neutron scattering \cite{klingler2018spin,li2020coherent,fan2020manipulation,quarterman2022probing}. This is due to the oxygen-mediated superexchange coupling mechanism between the Fe-O bond in YIG and the metallic atoms (Fe or Co) in the adjacent FM layer \cite{GrutterPRM22,QianPRM24}. Nevertheless, a clean interfacial superexchange coupling would require careful processing of the surface, such as ion milling \cite{li2020coherent,QianPRM24} or acidic treatments, which has been shown to improve the interfacial \textcolor{black}{spin pumping efficiency} \cite{JungfleischAPL13}. Such AFM coupling has been deployed for several magnon manipulation strategies: e.g., Fan \textit{et al}. \cite{fan2020manipulation} demonstrated a magnon spin valve by exploiting antiparallel magnetization configurations to suppress magnon transmission. Chen \textit{et al}. \cite{chen2018strong} reported an enhanced magnon-magnon coupling in antiparallel (AP) Co/YIG bilayers compared to parallel (P) states. 

In addition to the real part of the coupling ($J$), an imaginary component ($J'$) arises from spin pumping \cite{li2020coherent}, an effect which describes the flow of pure spin current \cite{TserkovnyakRMP05} between the two layers. This effect manifests as mutual damping enhancement (or reduction), depending on the symmetry of the hybrid magnon-magnon mode: the total linewidth increases for the out-of-phase hybrid mode and decreases for the in-phase mode in YIG/Py. In YIG/Py, $J'$ is measured to be $0.019\pm0.009$ mJ/m$^2$, which is about one third of $J$. This imaginary interfacial coupling works similarly as the dissipative coupling in cavity magnonics \cite{HarderPRL18}. However, since it usually comes with the much stronger real interfacial exchange coupling, the observed magnon-magnon coupling is dominated by the level repulsion (avoided crossing) rather than level attraction.

\paragraph{All-insulator system}

To mitigate high dissipation in metallic films, recent works have expanded to all-insulator heterostructures. In such systems, both layers can exhibit significantly lower magnetic damping than metallic FM films, resulting in a higher cooperativity. In addition, the interface quality is also greatly enhanced due to epitaxy growth. For example, Liu \textit{et al}. \cite{liu2024strong} achieved low-dissipation magnon-magnon coupling in epitaxial YIG/Tm$_3$Fe$_5$O$_{12}$(TmIG) bilayers, where interfacial exchange coupling enables coherent interactions at practical frequencies and wavevectors, overcoming the limitations of YIG/Py systems. The YIG/TmIG system exhibits antiferromagnetic interfacial exchange coupling, similar to YIG/metallic FM systems, with a coupling strength of $J\sim 0.06$ mJ/m$^2$. Similarly, Li \textit{et al}. \cite{li2024reconfigurable} demonstrated programmable magnonic crystals in epitaxial YIG/Gd$_3$Fe$_5$O$_{12}$(GdIG) by leveraging temperature-dependent \textcolor{black}{Gd$^{3+}$} moments to switch between AFM and FM coupling, at the magnetization compensation temperature of GdIG around 200 K. The coupling strength is $J\sim 0.5$ mJ/m$^2$ above 200 K, but it abruptly switches to $-0.5$ mJ/m$^2$ below 200 K and further increases from $-0.5$ mJ/m$^2$ to $-1.7$ mJ/m$^2$ as the temperature decreased to 50 K. These advances highlight ferrimagnetic insulators as versatile platforms for quantum magnonics.

\subsection*{\textcolor{black}{Weakly-coupled} sublattices in antiferromagnets}

Antiferromagnets (AFMs) consist of spin sublattices with antiparallel alignment that fully compensate the total magnetization in the materials. Like other multilayer ferromagnets, the spin dynamics in AFMs exhibit acoustic modes and optical modes, which corresponds to the in-phase and out-of-phase dynamics of different spin sublattices. Nevertheless, because of the antiparallel spin alignment, the eigenfrequencies of the acoustic and optic modes formed by different spin sublattices evolve differently with an external magnetic field, leading to a reduction of the higher-frequency optical mode and an increase of the lower-frequency acoustic mode and, therefore, their mode intersection at a certain field. Under certain conditions which break the symmetry and lift the orthogonality of the acoustic and optical modes such as rotating the biasing field to off-axis directions, the two modes will couple to each other, yielding their mode anticrossing or magnon-magnon mode hybridization.

The recent discovery of 2D AFM families adds additional versatility to studying GHz-range AFM magnon-magnon coupling, because of the weak interlayer exchange coupling strength. Symmetry analysis of different AFM magnon modes is demonstrated to be an easy and effective method to \textcolor{black}{predict if certain couplings exist or vanish, and thus can guide further experimental design.} With the development of THz spectroscopy and ultra-fast laser technology, rich nonlinear effects have been uncovered in bulk AFM materials, enabling exotic magnon mode operations including frequency up/down-conversion and second-harmonic generation. In the following section, we introduce the physical mechanisms and experimental results on the exchange-coupled magnons in AFM and 2D systems.

\begin{figure*}[htb]
\centering
\includegraphics[width=6.7 in]{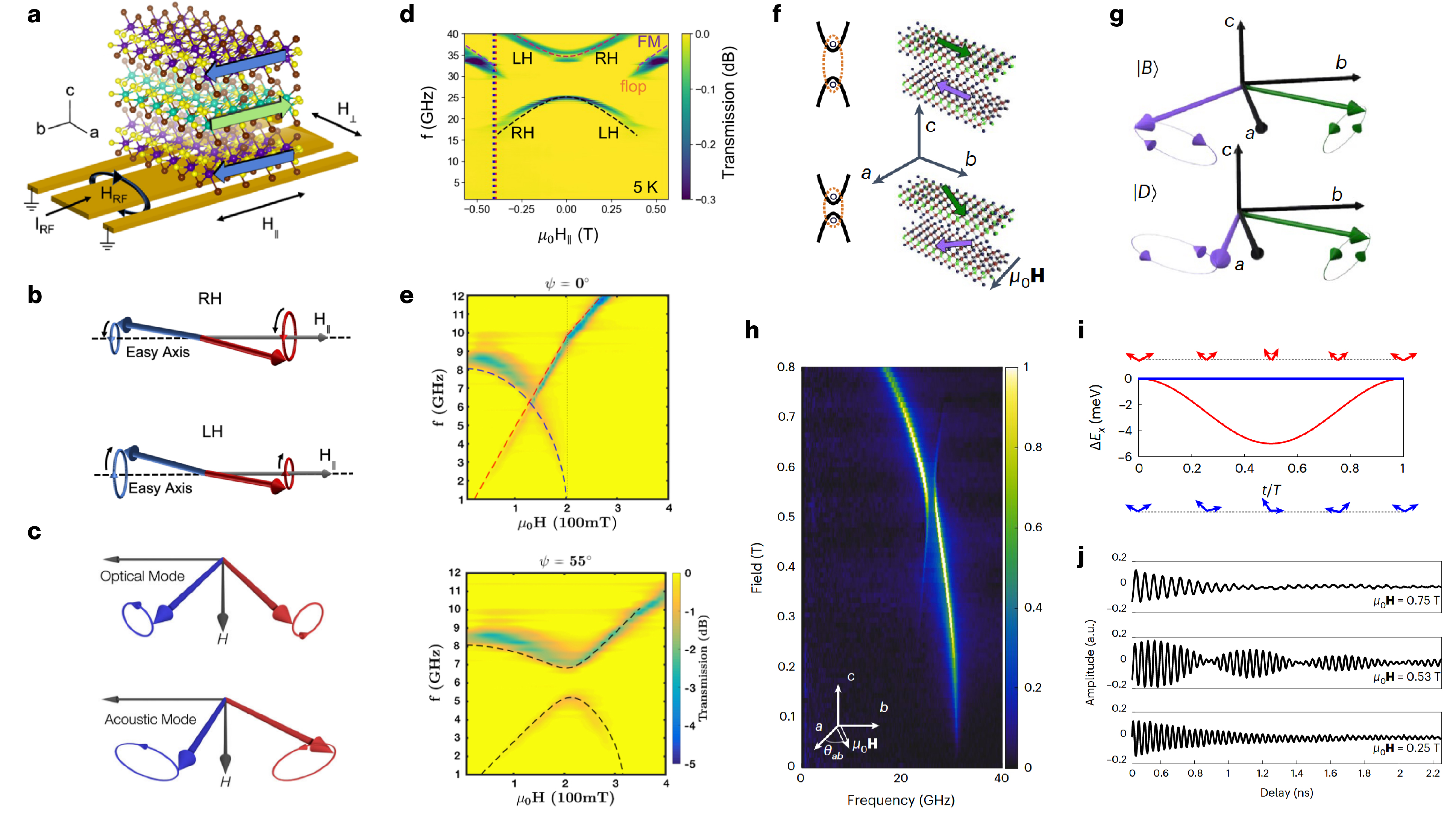}
\caption{Detecting magnon-magnon coupling in 2D AFM systems. (a) Schematic of microwave absorption experiment with 2D material. (b) Right-handed (RH) and left-handed (LH) magnon modes. (c) Optical and acoustic magnon modes. (d) Strong magnon-magnon coupling between RH and LH modes in CrSBr. (e) Strong magnon-magnon coupling between optical and acoustic modes in CrCl$_3$. The coupling is only activated by breaking in-plane rotational symmetry. (f) Magnetic state alters the exciton resonance energy in CrSBr. (g) Bright and dark magnon modes. (h) Magnon dispersion detected with optical reflectivity measurements in CrSBr. (i) Calculated transient exciton resonance energy shift from the optical (red) and acoustic (blue) magnon modes. (j) Coherent magnon hybridization between bright and dark magnon modes detected by time-resolved measurement. (Figures (a-d) taken from Ref. \cite{cham2022anisotropic}; (e) taken from Ref. \cite{macneill2019gigahertz}; (f-j) taken from Ref. \cite{diederich2023tunable}.)}
\label{fig_2D1}
\end{figure*} 

\paragraph{2D AFM}
The advent of 2D AFM materials has opened new avenues for studying magnon coupling in atomically thin systems. MacNeill \textit{et al}. \cite{macneill2019gigahertz} and Cham \textit{et al}. \cite{cham2022anisotropic} employed FMR to probe magnon interactions in CrCl$_3$ and CrSBr, respectively [Fig.\ref{fig_2D1} (a-e)]. While both materials share a Hamiltonian with in-plane FM and interlayer AFM couplings, their anisotropy profiles differ dramatically: CrCl$_3$ exhibits negligible in-plane anisotropy, favoring spin-flop states under any in-plane field, whereas CrSBr’s strong uniaxial anisotropy stabilizes antiparallel spin alignment below a critical field. \textcolor{black}{For CrCl$_3$, breaking the two-fold rotational symmetry of the coupled LLG equations around the in-plane field direction via an out-of-plane magnetic field component activates coupling between acoustic and optical magnon modes. In contrast, the same two-fold rotational symmetry around the in-plane easy axis of CrSBr can be broken by rotating the field in plane, creating an avoided crossing for acoustic and optical modes.} Cham \textit{et al}. \cite{cham2022anisotropic} further observed chirality-selective coupling between left-handed (LH) and right-handed (RH) magnons, with triaxial anisotropy inducing a zero-field avoided crossing—a hallmark of strong mode hybridization. Most recently, Li \textit{et al}. \cite{li2023ultrastrong} reported ultrastrong coupling (\textcolor{black}{normalized coupling rate $\eta_{\text{eff}} = g_{\text{eff}} /(2\pi f_\text{r}) = 0.31$ exceeding 0.1}) in CrPS$_4$, attributed to magnetocrystalline anisotropy rather than exchange enhancement, alongside chirality-switchable sublattice magnons—a feature also tied to the orthorhombic anisotropy. In parallel, Cham \textit{et al}. \cite{cham2025Science} recently demonstrated electrically tunable antiferromagnetic resonance in bilayer 2D AFMs using a spin-filter tunneling geometry, achieving both local detection and electrical control of magnon modes via spin-orbit torque–mediated damping modulation. This represents a major step toward integrating high-frequency AFM dynamics into nanoscale devices and leveraging sublattice-specific spin control for on-chip applications.

Despite these advances, significant challenges remains in probing 2D or bulk AFM magnon modes due to relatively weak FMR signals and their extension into the THz-frequency regime. These properties limit the effectiveness of conventional microwave techniques operating in the GHz range, which typically rely on magnetic susceptibility coupling. To address this, innovative optical approaches have emerged. For instance, Bae \textit{et al}., \cite{bae2022exciton} developed an exciton-mediated pump-probe reflectivity method to detect magnon dynamics in 2D AFM CrSBr, where coherent magnon dynamics is directly observed. In this scheme, spectral shifts of exciton resonances act as a sensitive probe for spin precession, effectively serving as a magnon proxy [Fig.\ref{fig_2D1} (f-j)]. The exciton resonance couples strongly to interlayer spin alignment, functioning as both a readout mechanism and a stabilizer for long-lived magnon modes. By tuning external magnetic fields and mechanical strain, precise control of opto-mechanical-magnonic coupling has been achieved \cite{diederich2023tunable}. With the improvement in measurement sensitivity, future study may uncover the unique layer-dependent magnon dynamics in 2D AFM systems.

\begin{figure*}[htb]
\centering
\includegraphics[width=6.7 in]{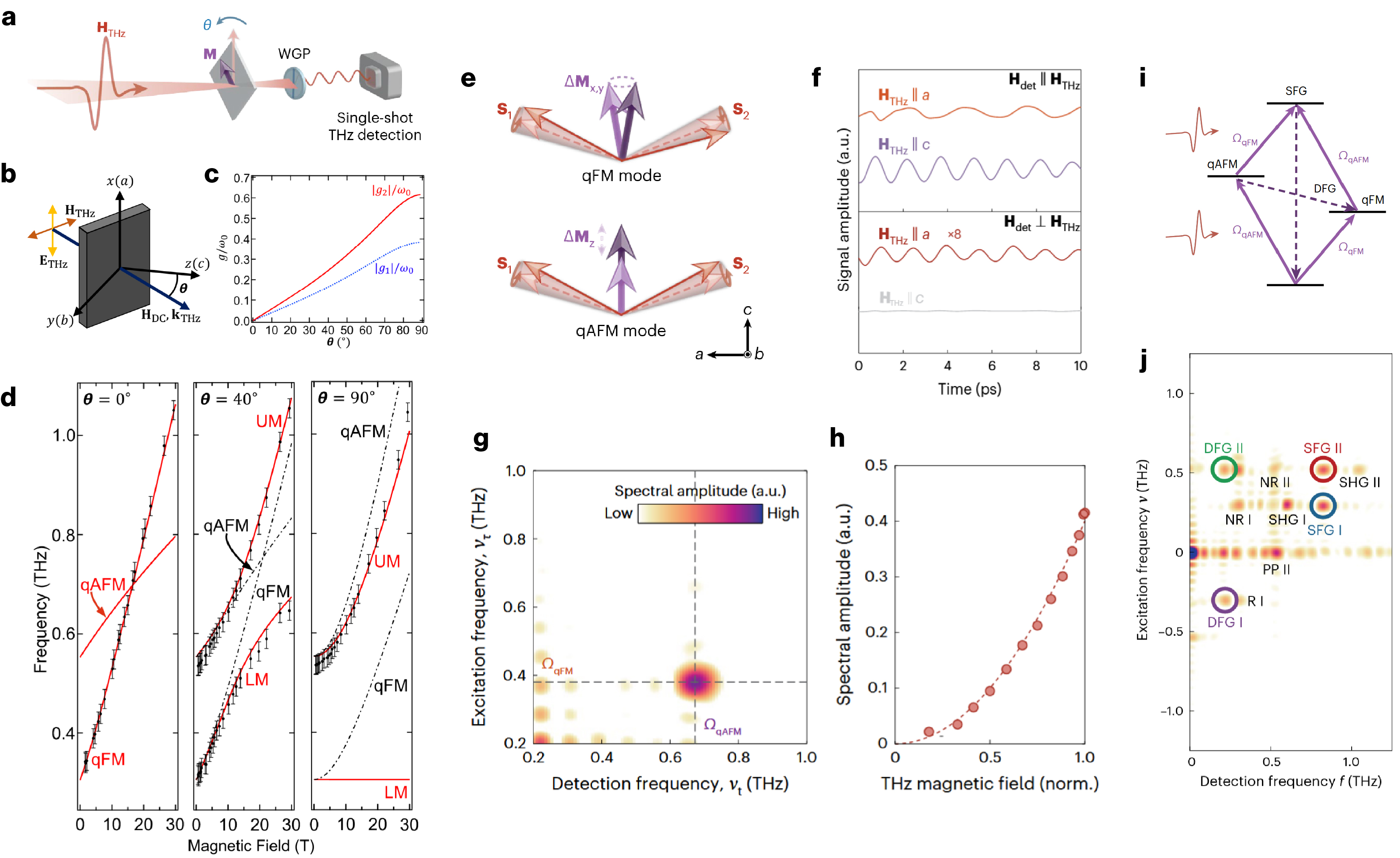}
\caption{Detecting magnon-magnon coupling in bulk AFM systems. (a) Schematic of time-resolved THz free induction decay (FID) measurement. (b) Sample geometry of YFeO$_3$ in a tilted magnetic field to study the THz frequency magnon bands. (c) Calculated co-rotating (blue) and counter-rotating (red) coupling strengths show the dominance of the counter-rotating coupling terms. (d) Measured THz magnon bands and the ultra-strong magnon-magnon coupling in YFeO$_3$. (e) quasi-FM (qFM) and quasi-AFM (qAFM) magnon mode. (f) Excitation of different magnon modes in ErFeO$_3$. (g) Strong magnon frequency up-conversion illustrated in 2D THz FID spectra. (h) Dependence of the magnon up-conversion signal amplitude on the pump magnetic field. (i) Magnon energy-level diagrams. (j) The 2D THz spectra in YFeO$_3$ summarizing the rich linear and nonlinear processes including pump-probe (PP),  rephasing or photon echo (R), non-rephasing (NR), two-quantum (2Q), second harmonic generation (SHG), sum-frequency generation (SFG) and difference frequency generation (DFG). (Figures (a),(e),(i),(j) taken from Ref. \cite{zhang2024terahertznonlinear}; (b-d) taken from Ref. \cite{makihara2021ultrastrong}; (f-h) taken from Ref. \cite{zhang2024terahertzupconversion}.)}
\label{fig_2D2}
\end{figure*} 

\paragraph{Bulk AFM}
Using time-domain two-dimensional THz spectroscopy, Zhang \textit{et al}. \cite{zhang2024terahertzupconversion,zhang2024terahertznonlinear} investigated the 3D canted antiferromagnets ErFeO$_3$ and YFeO$_3$ [Fig.\ref{fig_2D2} \textcolor{black}{(a,g-j)}]. Under weak perturbations, both materials exhibit two magnon modes: the quasi-ferromagnetic (qFM) mode, corresponding to a precession of the magnetization orientation, and the quasi-antiferromagnetic (qAFM) mode, representing a periodic modulation of the magnetization amplitude \textcolor{black}{[Fig.\ref{fig_2D2} (e)]}. Crucially, both ErFeO$_3$ and YFeO$_3$ possess Dzyaloshinskii–Moriya interactions that lead to canted magnetic moments. These canted moments break time-reversal symmetry and enable magnetic dipole–based nonlinear interactions \cite{fiebig1994second}. By applying two-dimensional time-resolved THz spectroscopy, Zhang \textit{et al}. were able to selectively and directly probe various types of coherent nonlinear magnon–magnon interactions in YFeO$_3$ \textcolor{black}{[Fig.\ref{fig_2D2} (f)]}. Although the linear magnon response dominates in conventional one-dimensional THz spectroscopy, the 2D technique revealed clear signatures of anharmonic magnon interactions in off-diagonal peaks, which arise from coherent interactions between the qAFM and qFM modes. Notably, second-order magnon processes—such as sum-frequency generation, second-harmonic generation, and difference-frequency generation—were observed for the first time \textcolor{black}{[Fig.\ref{fig_2D2} (j)]}. Using a similar approach, the authors also demonstrated asymmetric up-conversion processes in ErFeO$_3$ \textcolor{black}{[Fig.\ref{fig_2D2} (g)]}. It is important to emphasize that while strong THz-field driving is essential, the presence of two coupled resonant magnon modes (magnon–magnon coupling) is equally critical for enabling these asymmetric up-conversion processes.

\subsection*{Syn-AFMs (RKKY-driven)}

The RKKY interaction, a fundamental phenomenon in magnetism, plays a pivotal role in a variety of magnonic and spintronic applications. The RKKY interaction arises from the indirect coupling of spins in magnetic materials via conduction electrons. In this section, we summarize recent progress that has been made in understanding the nature of magnon-magnon interactions within magnetic heterostructures which utilize the RKKY interaction.  

\begin{figure*}[htb]
\centering
\includegraphics[width=6.5 in]{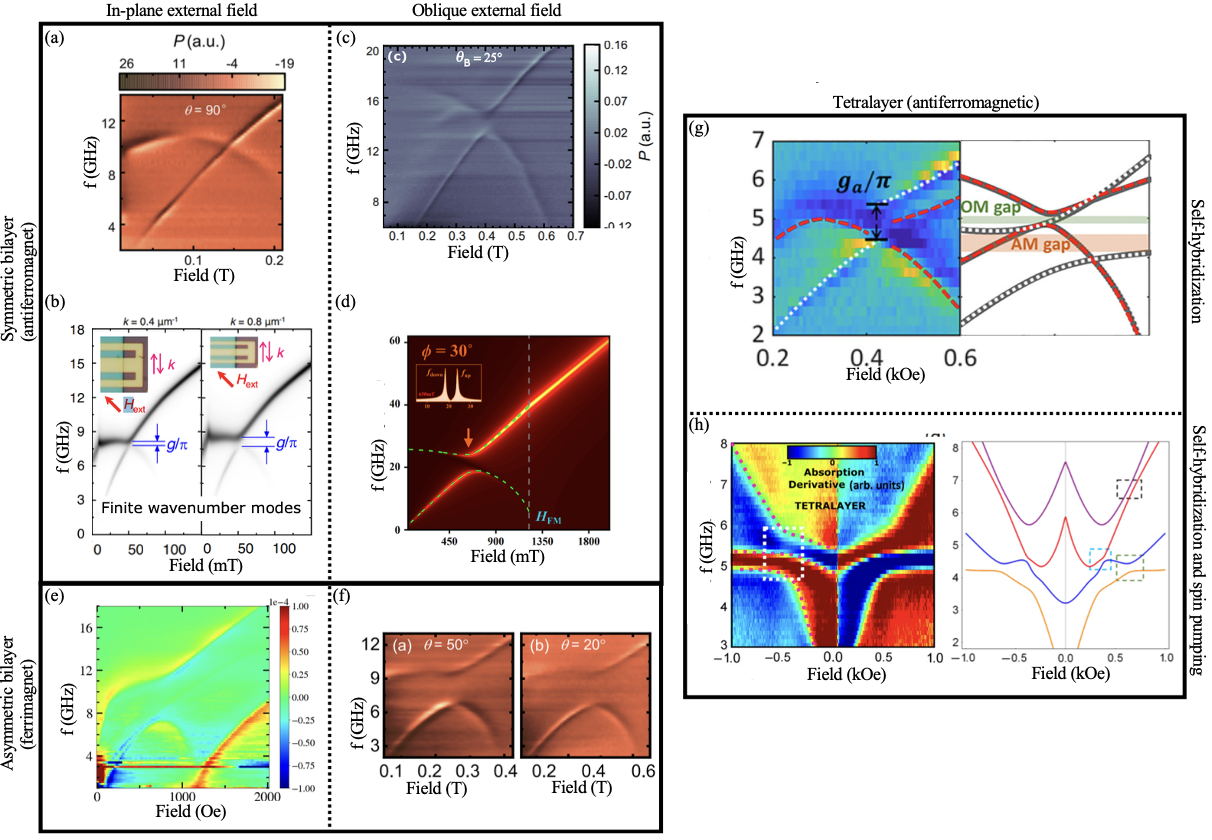}
\caption{Interactions between acoustic and optical magnons in synthetic magnets are summarized above. The upper left portion of the figure illustrates how, for synthetic antiferromagnets, the crossing between an acoustic and optical branch [(a)] can turn into an avoided crossing by changing the wave number [(b)], or applying a symmetry breaking external field oblique to the sample plane, [(c), (d)].  The respective references for these images are taken from \cite{sud2023magnon}, \cite{shiota2020tunable}, \cite{sud2020tunable}, and \cite{dai2021strong}. Similar avoided crossings are shown in the lower left portion of the figure in a synthetic ferrimagnet, where the magnetic layers have unequal thicknesses or dissimilar magnetizations, [(e), (f)] taken from Ref. \cite{hossain2024broken,sud2023magnon}. An alternative strateegy to engineer interactions into these materials is by changing the number of magnetic layers. The right portion of the figure, [(g), (h)] taken from Ref. \cite{rong2024layer,subedi2025engineering}, shows how in a tetralayer, acoustic or optical magnon pairs interact. If interlayer spin pumping is present in the tetralayer, optical and acoustic magnon pairs can interact as well without the aid of a symmetry breaking field.}
\label{fig_saf_sum}
\end{figure*} 

\paragraph{Metallic multilayer syn-AFMs}
In multilayer heterostructures, comprised of magnetic thin films separated by nonmagnetic spacer layers, the magnetic films can be exchange coupled by the RKKY interaction. The best known structure in this category is the syn-AFMs. A conventional syn-AFM structure consists of two magnetic thin films separated/coupled together via one spacer layer.  Often, the only anisotropy within the magnetic layers is from the shape of the magnetic layers, e.g., many syn-AFMs  behave like easy-plane antiferromagnets. In an easy-plane antiferromagnet, there are two antiferromagnetic resonance (AFMR, $k = 0$) magnon modes that can be excited with a uniform driving field. The so-called acoustic AFMR mode has a zero frequency at zero field, and increases linearly with the applied external field. The so-called optical AFMR mode has a finite frequency at zero field, which is proportional to the square root of the product of the interlayer exchange field and the magnetization of the individual layers. The frequency of the optical AFMR mode decreases as the external magnetic field increases. At some finite external field, these two modes will encounter energy degeneracy, and there will be a crossing point. The magnon-magnon interactions that are studied in these RKKY coupled materials most often tend to hybridize the acoustic and optical AFMR modes in the vicinity of this crossing point. Consequently, these interactions manifest as an avoided energy level crossing between the two modes of interest.

Before delving into the origins of the magnon-magnon interactions in syn-AFMs, it is important to first describe the role that the RKKY interaction plays in dictating the general dynamic properties of a syn-AFM. Importantly, the strength of the interlayer RKKY exchange interaction greatly impacts the external field and frequency where the acoustic and optical magnon modes cross. So, although the RKKY interaction is not necessarily responsible for the magnon-magnon interaction within a syn-AFM, it is responsible for the presence of both optical and acoustic modes as well as the field-frequency pairs where the modes interact. Therefore, if one is motivated to investigate magnon-magnon interactions at higher frequencies, it is important to optimize the strength of the RKKY interaction itself. In recent years, the most commonly used spacer layer is ruthenium. Many of the highest zero-field optical magnon frequencies, above 20 GHz, have been reported in CoFeB/Ru systems \cite{waring2020zero,zhou2022self,mouhoub2023exchange}. More recently, in syn-AFM structures comprised of Co and Ru, the antiferromagnetic interlayer exchange interaction was further enhanced by alloying the Ru spacer layer with magnetic materials, e.g., Co and Fe \cite{winther2024antiferromagnetic}.

\paragraph{Magnonic Interactions in syn-AFMs}
Following the report of tunable magnon-magnon interaction in layered CrCl$_3$ crystals in 2019 \cite{macneill2019gigahertz}, the same tunable interaction was demonstrated by Sud \textit{et al}., in an syn-AFM with easy-plane anisotropy \cite{sud2020tunable}. This interaction, between the uniform ($k$ = 0) acoustic and optical magnon modes, is controlled by tilting the external field out of the easy plane, as shown in \textcolor{black}{Fig.} \ref{fig_saf_sum}(c) and (d). The origin of this interaction is from the external field breaking \textcolor{black}{the two-fold rotational symmetry of the system, as explained for 2D AFM materials, which in general exists in any easy-plane antiferromagnet \cite{macneill2019gigahertz, sud2020tunable, li2021symmetry, dai2021strong}}. This symmetry can also be broken by intrinsic aspects of a syn-AFM. He \textit{et al}. demonstrated that weak uniaxial anisotropy could be induced in systems like CoFeB/Ir/CoFeB. In this system, the magnon mode gap could be tuned based on the orientation of the external field relative to the easy and hard axes \cite{he2021anisotropic}. \textcolor{black}{Similarly, Wang \textit{et al}. theoretically and experimentally} explored how \textcolor{black}{asymmetric magnetic anisotropies of the two ferromagnetic layers in a FM/NM/FM structured syn-AFM} can be used in lieu of the out-of-plane field to hybridize the uniform acoustic and optical magnon mode \cite{wang2024ultrastrong}. 

An alternative way to hybridize acoustic and optical magnons in easy-plane \textcolor{black}{syn-AFM} was demonstrated by Shiota \textit{et al}. \cite{shiota2020tunable}, [Fig. \ref{fig_saf_sum} (b)]. Their work showed that by adjusting the in-plane magnetic field direction relative to the wave vector of the excited magnon modes, they could modulate the coupling strength, with stronger coupling observed at higher wave numbers.  In these experiments, no out-of-plane external field was required; the magnon-magnon interaction originated from dynamic dipolar fields generated between the layers when the wave number was non-zero.    

It is also possible to engineer magnon-magnon interactions into syn-AFMs, without the aid of symmetry breaking external fields, through the use of layer number-dependent effects. For example, by altering the structure of a syn-AFM to have four layers the AFMR spectra \textcolor{black}{are} altered to have two optical and two acoustic modes.  The pairs of all-optical, or all-acoustic modes ``self-hybridize'' with each other without the aid of a symmetry breaking external field \cite{sklenar2021self}. This self-hybridization manifests as characteristic avoided energy level crossing between the acoustic or optical pair of modes.  Micromagnetic simulations have been used to demonstrate the potential for electrically tuning this type of magnon-magnon interaction within a tetralayer \cite{sklenar2021self}. 
Rong \textit{et al}. \textcolor{black}{recently} demonstrated these effects, and observed how the presence or absence of magnon-magnon interactions was strongly dependent on the number of layers in syn-AFMs \cite{rong2024layer}.  In particular, they observed magnon-magnon interactions arising from self-hybridization within the even-layered syn-AFMs, as shown in \textcolor{black}{Fig.} \ref{fig_saf_sum}(g), while odd-layered syn-AFMs exhibited avoided energy level crossings due to the structural asymmetry within the syn-AFM. In similar tetralayer structures, Subedi \textit{et al}. also observed an avoided energy level crossing between optical and acoustic magnons, as shown in \textcolor{black}{Fig.} \ref{fig_saf_sum}(h), which could not be attributed to the self-hybridization effect.  \textcolor{black}{Here, the magnon-magnon interaction arises from symmetry breaking which in this case is amplified by interlayer dynamic field-like torques originated through spin pumping between the interior and surface layers of the structures \cite{subedi2025engineering}. When small structural asymmetries (interfacial roughness) exist within a syn-AFM, the additional dynamic field-like torques, from spin pumping, generate a symmetry breaking field that is strong enough to clearly hybridize acoustic and optical magnons.}

\paragraph{Magnonic Interactions in synthetic Ferro- and Ferri-magnets}
In addition to synthetic antiferromagnets, one can create both synthetic ferrimagnets and ferromagnets. A synthetic ferrimagnet can be made by modifying a conventional synthetic antiferromagnet such that the net magnetic moment in each magnetic layer (or sub-lattice) is different. For example, ferrimagnetic behavior can be observed in structures where magnetic layers have unequal thicknesses, dissimilar ferromagnetic materials, or an odd number of magnetic layers \cite{sud2023magnon,subedi2024even, subedi2025investigating, hossain2024broken}. Magnon-magnon interactions in synthetic ferrimagnets, where the magnetic layers have different thicknesses and/or the magnetic layers are comprised of different magnetic materials, have been explored \cite{sud2023magnon,hossain2024broken}, as shown in \textcolor{black}{Fig.}\ref{fig_saf_sum}(e) and (f). In the case of a synthetic ferrimagnet, with two magnetic layers, there is an optical and acoustic ferrimagnetic resonance mode akin to the modes found in a syn-AFM. In a synthetic ferrimagnet, these modes naturally hybridize since the coupled LLG equations are intrinsically not symmetric under the symmetry operation where the magnetization of each layer is rotated around the external field by 180$^\circ$ and then followed by a layer swap.

In RKKY coupled systems, synthetic ferromagnets are materials where the magnetic thin films are separated by a spacer layer which promotes a ferromagnetic interlayer exchange coupling. In these materials, it is generally more difficult to characterize the strength of the RKKY interaction and to excite any magnon mode aside from the uniform (acoustic) mode. In general, characterization of the RKKY interaction is difficult due to the tendency for the two layers to be aligned. Magnetometry cannot be used to estimate the interlayer exchange interaction between two magnetic thin films unless the magnetization is non-collinear. The excitation of optical magnon modes in a synthetic ferromagnet is difficult because it requires a spatially non-uniform driving field. Conventional FMR-based techniques typically generate spatially uniform fields that can only couple to the acoustic mode, which behaves like a conventional Kittel-like mode even in the presence of the interlayer coupling. These challenges, perhaps, underpin the relative lack of experimental progress in studying magnon-magnon interactions with synthetic ferromagnetic materials.

Another class of synthetic magnets, with magnonic properties that have not been well explored experimentally, is comprised of magnetic layers with orthogonal anisotropies or tilted anisotropies. Such structures have been theoretically modeled using the Landau Lifshitz equations \cite{tengfei2025magnon} as well as micromagnetic simulations \cite{chen2025micromagnetic}. In the case of two magnetic layers with orthogonal anisotropies, the coupling between acoustic and optical magnon modes was investigated as both a function of the interlayer exchange interaction as well as the saturation magnetization. In the case of two magnetic layers with tilted anisotropy, the interplay of  \textit{both} the external field tilt and the anisotropy tilt on the interaction strength between the acoustic and optical magnon modes were investigated. Synthetic magnets with orthogonal anisotropies are also a promising material platform because they have recently been shown to house interlayer antisymmetric exchange interactions that oscillate as a function of the spacer layer thickness, which strongly indicates an RKKY origin \cite{huang2025experimental}.

\subsection*{Dipolar coupling}

	The generation, manipulation, and detection of spin waves at the nanoscale are three key research directions in Magnonics. Recently, a significant mechanism for spin-wave manipulation at nanoscale has emerged—dynamic dipolar coupling, which complements static dipolar coupling.

\begin{figure*}[htp]
	\centering
	\includegraphics[width=0.95\textwidth]{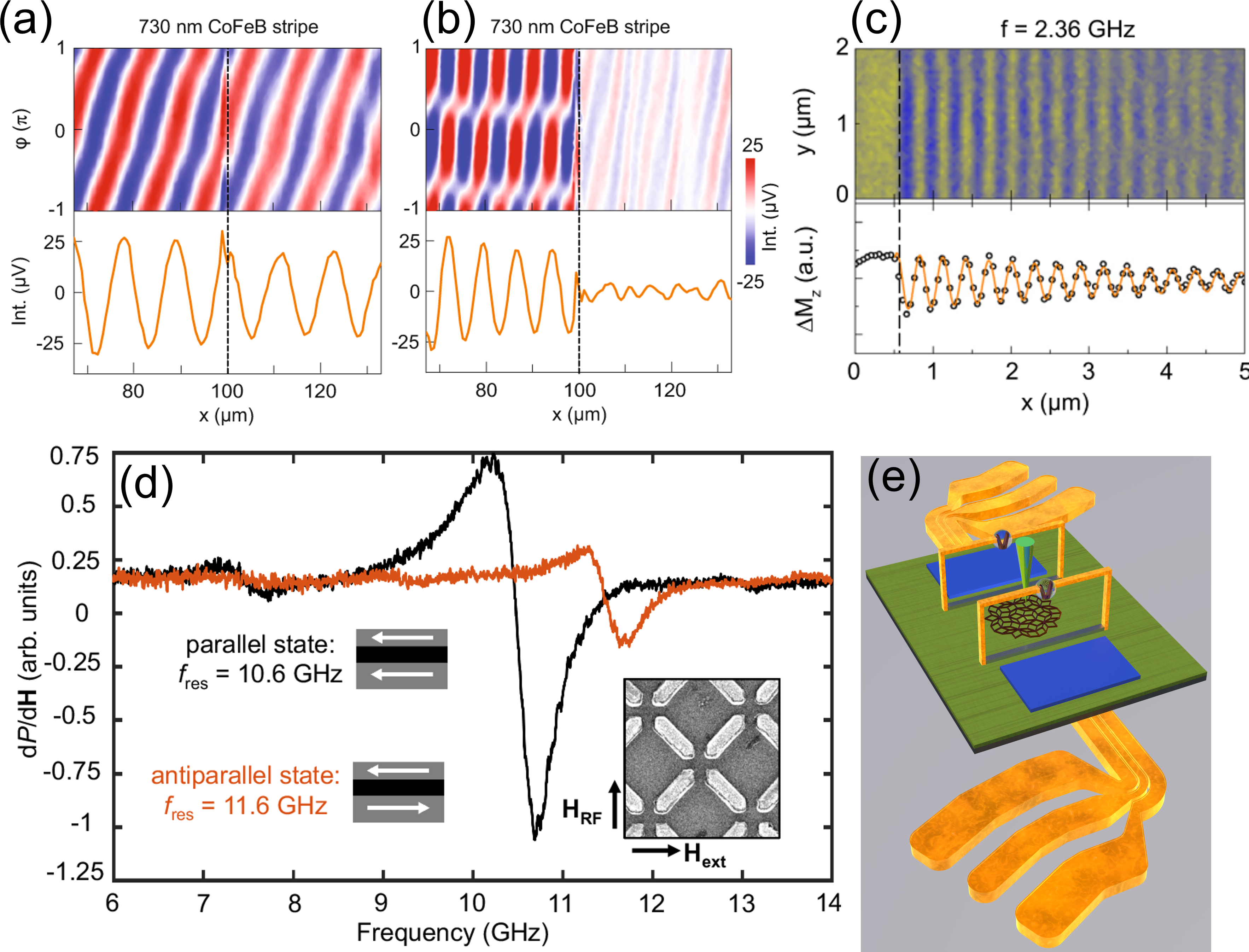}
	\caption{ 
		(a), (b) Phase-resolved TR-MOKE microscopy maps and line profiles measured on a 100-nm-thick Y\textsubscript{3}Fe\textsubscript{5}O\textsubscript{12} (YIG) film with a 730-nm-wide CoFeB stripe. Figure taken from ref. \cite{Qin2021}. 
		(c) High-resolution normalized TR-STXM images and line profiles of propagating short-wavelength spin waves in the YIG/Co bilayer at 2.36 GHz with a 10 mT bias field. Figure taken from ref. \cite{Talapatra2023}. 
		(d) Ferromagnetic resonance (FMR) spectra for parallel (black) and antiparallel (red) macrospin states, showing a 1 GHz frequency shift between the states at $H_{\text{ext}} = 0$. 
		An SEM schematic defines the DC external field $H_{\text{ext}}$ and RF field $H_{\text{RF}}$ orientations for the FMR data panels in this figure. Figure taken from ref. \cite{dion2024ultrastrong}.   
		(e) Schematic of the proposed all-on-chip device inspired by previous studies on dynamic dipolar coupling. 
		The Ni\textsubscript{81}Fe\textsubscript{19} (blue), Penrose P3 nanomagnet arrays (maroon), and Pt (gray) are patterned on plain YIG (dark green). 
		The entire chip is placed on an RF co-planar waveguide (CPW). A global spin-pumping voltage can be determined by connecting a nanovoltmeter to the Pt stripes, whereas the local mode profile can be obtained using BLS methods (bright green indicating the laser). 
	}
	\label{Fig1}
\end{figure*}	
	
	Dynamic dipolar coupling occurs when oscillating magnetic moments in one material (e.g., a Ni\textsubscript{81}Fe\textsubscript{19}
	 ferromagnetic stripe) or one element in an array generate time-varying magnetic fields that interact with moments in another material (e.g., YIG) or another element of the array. This interaction alters the spin-wave dispersion relation by modifying the effective magnetic field. \textcolor{black}{The static dipolar field induced by the first material or element \textcolor{black}{shifts the entire spin-wave dispersion in the second material or element, while} the dynamic dipolar field \textcolor{black}{created by the first element} induces changes within a narrow frequency range near resonance \textcolor{black}{in the second layer}. This offers} additional functionality including magnon-magnon \textcolor{black}{coupling} and hybridization \cite{Gubbiotti_2018,lendinezAppl.Phys.Lett.2021a,gartsideNatureCommunications2021,dion2024ultrastrong,montoncelloJ.Appl.Phys.2023,negrelloAPLMater.2022b}. This coupling is strongest when materials are in close proximity, separated only by a thin nonmagnetic spacer, and is highly dependent on their relative magnetization orientations.
	
	
	\textcolor{black}{One key effect of dynamic dipolar coupling is the downshift of the spin-wave dispersion relation in a continuous YIG film caused by dynamic dipolar coupling to a ferromagnetic metal nanostripe \cite{Qin2021}}, enabling mode conversion and nanoscale control. For instance, time-resolved magneto-optical Kerr effect (TR-MOKE) microscopy in a YIG/CoFeB bilayer shows that this coupling induces a frequency downshift, reducing spin-wave wavelengths from 12.8 $\mu$m to 310 nm at 1.76 GHz \cite{Qin2021} [Fig. \ref{Fig1}(a) and (b)]. Similarly, Talapatra et al. observed wavelength reductions from 3.4–7.5 $\mu$m to 280–480 nm at the YIG/Co interface, as seen in time-resolved scanning transmission x-ray microscopy (TR-STXM) images \cite{Talapatra2023} [Fig. \ref{Fig1}(c)].
	
	Beyond dispersion control, dynamic dipolar coupling facilitates reconfigurable spin-wave transport. It has been used in hybrid YIG-based material structures that function as Fabry-Pérot nano-resonators \cite{Qin2021}, where overlapping spin waves interfere constructively (in-phase) or destructively (out-of-phase), analogous to optical interference. The resonance condition in these structures depends on path length, wavelength, and phase shifts at interfaces or boundaries. Complementing these studies, Santos \textit{et al}. demonstrated an order-of-magnitude increase in magnon confinement in a YIG film region not covered by the ferromagnetic metal, preserving YIG’s optimal magnetic properties within the cavity \cite{Santos2023b}. Furthermore, the fabrication of micrometer-sized YIG cavities—created between two YIG/Permalloy bilayers— in an on-chip formulation offers a novel approach for coherent magnon control where a nanometer-sized Pt strip has been shown to function as a noninvasive local detector of magnon resonance intensity via spin pumping \cite{Santos2023b}.
	
	
	Nonreciprocity—where spin-wave propagation differs depending on direction—enhances magnonic functionality. Qin \textit{et al}. achieved partial and full nonreciprocity in YIG hybrid structures, with narrow FM stripes (730 nm width) exhibiting asymmetric dispersion and wider FM stripes (\textgreater 25 $\mu$m width) showing damping-related differences, respectively \cite{Qin2021}.
	
	Chiral scattering of spin waves, driven by dynamic dipolar coupling, has also been investigated. Fripp \textit{et al}. used micromagnetic simulations and a phenomenological model to explore this effect in a ferromagnetic thin film coupled to a nanoscale magnonic resonator  \cite{Fripp2021}. They demonstrated strong, nonreciprocal scattering, where the resonator’s precession induces direction-dependent effects via quasi-uniform (13.5 GHz) and dark (17.4 GHz) modes. This coupling, termed chiral due to its handedness in relation to spin-wave momentum, enables the development of magnonic diodes and phase shifters, offering promising avenues for scalable spin-wave control.
	
	
	Recent studies have extended the exploration of dynamic dipolar coupling to artificial spin lattices. A notable example is a dipolarly coupled trilayer system composed of Ni\textsubscript{81}Fe\textsubscript{19} (30 nm)/Al (35 nm)/Ni\textsubscript{81}Fe\textsubscript{19} (20 nm) nanobars arranged in a square lattice \cite{dion2024ultrastrong,bhat2025magnon}. The thicker nanobars were found to switch at higher fields compared to the 20 nm Ni\textsubscript{81}Fe\textsubscript{19} layers. Spin-wave spectroscopy revealed a nominal 1 GHz frequency shift between configurations where the top and bottom layers were anti-aligned versus aligned  \cite{dion2024ultrastrong} [Fig. \ref{Fig1}(d)]. Additionally, the application of external magnetic fields at different in-plane angles resulted in further reconfiguration of the spin-wave spectra  \cite{bhat2025magnon}.

    Another type of dynamically coupled systems emerges if artificial-ice lattices patterned on the top of continuous thin film underlayers. Negrello \textit{et al}. showed that an array of stadium-shaped NiFe nanoislands deposited on the top of a continuous NiFe film with non-magnetic spacer layers of varying thickness results in distinct modes in the film at either specific wavelengths or with intensity modulation imprinted by the artificial spin-ice system \cite{negrelloAPLMater.2022b}. This type of dynamic mode coupling in the vertical direction, facilitated by dipolar coupling, enables the modulation of spin-wave propagation at nanometer length scales \cite{negrelloAPLMater.2022b,montoncelloJ.Appl.Phys.2023}.

\textcolor{black}{Besides in magnetic multilayers, the dynamic dipolar coupling} can also arise in arrays of magnetic micro- and nanostructures if the constituent elements of the array are sufficiently close to one another \cite{adhikari2020large,adhikari2021observation,Adhikari_2021,gartsideNatureCommunications2021,lendinezAppl.Phys.Lett.2021a,bhat2021tuning}. For example, strong magnon-magnon coupling was observed in crossshaped nanoring arrays, nanocross arrays, and artificial spin ice systems. Interestingly, Adhikari and co-workers observed an enhancement of inter-element-dynamic dipolar interactions leading to \textcolor{black}{avoided level-crossing} and nonlinear frequency shifts \cite{Adhikari_2021}. 

The next step in this direction can involve integrating 3D nanostructures onto on-chip resonators and employing both global spin pumping and local probe techniques, such as Brillouin light scattering microscopy [Fig. \ref{Fig1}(e)]. The in-plane dipolar and exchange coupling can be further tuned by modifying disconnected and interconnected networks of nanobars on periodic and aperiodic \cite{bhat2023spin} lattices. Additionally, a systematic reduction in periodicity through controlled disorder  \cite{saccone2019towards} could provide new avenues for engineering magnonic behavior.
	
These advancements in dynamic dipolar coupling, nonreciprocity, and artificial spin lattices pave the way for scalable, reconfigurable magnonic devices with applications in wave-based computing, signal processing, and nanoscale information transport \cite{pal2024using,gubbiotti20252025}.

\begin{table*}[htb]
\begin{tabular}{lccccc}
{Systems} & {freq. (GHz)} & {Coupl. strength (GHz)} & {Magnon modes} & {Coupl. types} & {Ref.}\\
\bf{Direct-Ex} \scriptsize{(Magnet-1 / Magnet-2)}\\
YIG(100 nm)/NiFe(9 nm) & -- & 0.35 & PSSW/FMR & SC & \cite{li2020coherent}\\ 
YIG(3000 nm)/NiFe(10 nm) & 4-8 & 0.09 & PSSW/FMR & MIT & \cite{xiong2020probing,xiong2022tunable}\\
YIG(3000 nm)/NiFe(30 nm) & 2.5-3.5 & 0.022 & PSSW/FMR & MIT/SSC & \cite{inman2022hybrid}\\
YIG(100 nm)/NiFe(30 nm) & 1-9 & -- & BVSW/MSSW & MIT & \cite{santos2023magnon}\\
YIG(20 nm)/Co(30 nm)  & -- & 0.79 & PSSW/FMR & SC/USC & \cite{chen2018strong}\\
YIG(20 nm)/Ni(20 nm) & -- & 0.12 & PSSW/FMR & SC/USC & \cite{chen2018strong}\\ 
YIG(1000 nm)/Co(50 nm) & 0-25 & 0.2 & PSSW/FMR & SC & \cite{klingler2018spin}\\
YIG(295 nm)/CoFeB(50 nm) & 2-8 & -- & PSSW/FMR & SC & \cite{qin2018exchange}\\  
TmIG(350 nm)/CoFeB(50 nm) & -- & 0.263 & PSSW/FMR & SC & \cite{liu2024strong}\\ 
TmIG(140 nm)/YIG(140 nm) & -- & 0.105 & PSSW/FMR & SC & \cite{liu2024strong}\\ 
GdIG(74 nm)/YIG(46 nm) & 10-15 & 0.5(150 K) & PSSW/FMR & SC & \cite{li2024reconfigurable}\\ 
\bf{Direct-Ex} \scriptsize{(AFM Magnet)}\\
CrCl$_3$ (crystal) & 10-25 & 0-1.37(1.56 K) & Acoustic/Optic & SC & \cite{macneill2019gigahertz}\\ 
CrSBr (crystal) & 15-40 & 4(100 K) & Acoustic/Optic & SC/Chiral & \cite{cham2022anisotropic}\\ 
CrSBr (flake, 20 nm) & 20-30 & $\sim$ 8(35 K) & Acoustic/Optic & SC/Strain & \cite{diederich2023tunable}\\  
CrPS$_4$ (crystal) & 10-25 & $>$4(22.5 K) & Acoustic/Optic & USC/Chiral & \cite{li2023ultrastrong}\\  
YFeO$_3$ (crystal) & $\sim$ 600-1000 & $>$350 & qAFM/qFM & USC & \cite{makihara2021ultrastrong}\\ 

\bf{Dipolar}\\
YIG(70 nm)/CoFeB(stripe) & -- & -- & MSSW/BVSW & -- & \cite{qin2021nanoscale}\\ 
YIG(80 nm)/CoFeB(stripe) & 1-13 & -- & PSSW & -- & \cite{sheng2023nonlocal}\\
YIG(30 nm,PMA)/CoFeB(stripe) & 5-10 & -- & \textcolor{black}{FVSW} & MIT & \cite{wang2023reconfigurable}\\
YIG(20 nm)/Co(grating) & 15-20 & 0.62 & MSSW/PSSW & MIT & \cite{chen2019excitation,liu2018long}\\  
NiFe(44 nm, film) & 10-15 & $\sim$1 & MSSW/PSSW & SC & \cite{song2021nonreciprocal}\\
NiFe(20 nm)/YIG(3900 nm) & 10-15 & 0.5/1.3 & MSSW/PSSW & SC & \cite{kong2024dipolar}\\
NiFe(20 nm, cross array) & -- & -- & MSSW & MIT & \cite{adhikari2020large}\\ 
NiFe(30 nm, spin ice) & 5-11 & 3.275 & Acoustic/Optic & USC & \cite{dion2024ultrastrong}\\

\bf{RKKY} \scriptsize{(Magnet / Interlayer)}\\
CoFeB(15 nm)/Ru(0.6 nm) & 3-17 & 0-0.67 & Acoustic/Optic & SC & \cite{shiota2020tunable}\\ 
CoFeB(3 nm)/Ru(0.5 nm) & 5-20 & $\sim$1 & Acoustic/Optic & SC & \cite{sud2020tunable}\\ 
(Co/Ni)/Ir(0.6 nm) & 0-18 & $>$4 & Acoustic/Optic & USC/DSC & \cite{wang2024ultrastrong}\\ 
NiFe(3.1 nm)/Ru(1 nm) & 2-7 & 0.556 & Acoustic/Optic & SC & \cite{rong2024layer}\\ 
NiFe(5 nm)/Ru(1 nm) & 2-8 & -- & Acoustic/Optic & SC & \cite{subedi2023magnon}\\ 

\bf{DMI}\\ 
(CH$_3$CH$_2$NH$_3$)$_2$CuCl$_4$ & 2-5 & $>$0.27(2.5 K) & Acoustic/Optic & SC & \cite{comstock2023hybrid}\\

\hline  
\end{tabular}
\caption{Summary of the \textcolor{black}{current magnon-magnon material systems exhibiting different coupling frequencies, coupling strengths, involved magnon modes, and coupling types. Involved magnon modes -- PSSW: perpendicular standing spin wave; FMR: ferromagnetic resonance (Kittel mode); MSSW: magnetostatic spin wave; \textcolor{black}{FVSW}: forward volume spin wave; BVSW: backward volume spin wave; Acoustic(optic): in-phase(out-of-phase) coupled resonances in syn-AFMs; qAFM(qFM): AFM(FM)-coupled spin resonances in crystal AFM. Coupling types -- SC: strong coupling, defined as the coupling strength $g$ being greater than the dissipation rates of both magnon counterparts, $g > \gamma_{m1}, \gamma_{m2}$; MIT: magnetically-induced transparency, defined as the coupling strength being greater than the dissipation of one mode but less than the other, $\gamma_{m1}> g> \gamma_{m2}$; USC, ultrastrong coupling, defined as the coupling strength $g$ being comparable to (a fraction of) the bare mode frequencies of the uncoupled systems; DSC, deep strong coupling: defined as the coupling strength $g$ being greater than the bare mode frequencies of the uncoupled systems; SSC: superstrong coupling, defined as the coupling strength $g$ being comparable to the free-spectral range (FSR). } }   
 \label{Tab:list}
\end{table*}

\subsection*{Dzyaloshinskii–Moriya interaction (DMI)}

The DMI favors an orthogonal spin configuration. It can be described using a Hamiltonian given by, $\mathbf{H}_\text{DM} = -\mathbf{D}_{12} \cdot (\mathbf{S}_1 \times \mathbf{S}_2)$, where $\mathbf{D}_{12}$ is the DMI vector, $\mathbf{S}_1$ and $\mathbf{S}_2$ are two atomic spins. The competition between the DMI and other types of spin-spin interactions, such as the Heisenberg exchange interaction, which favors a collinear spin configuration, enables the stabilization of a diverse range of unconventional spin textures (e.g., skyrmions). The presence of a large DMI simultaneously requires a broken inversion symmetry and strong spin-orbit coupling (SOC) in the framework of the three-site model proposed by Fert and Levy \cite{FertLevyPRL1981,FertNatNano2013}. The breaking of inversion symmetry can occur naturally in lattices of crystals \cite{TokuraChemRev2020}, such as B20 metals (e.g., $\mathrm{Fe}_{\mathrm{1-x}}\mathrm{Co}_{\mathrm{x}}\mathrm{Si}$ \cite{XZYuNature2010FeCoSi}, FeGe \cite{XZYuNatMaterFeGe2010}, MnSi \cite{MnSiSkrymionNL2012}) and insulators with similar crystal structure (e.g., $\mathrm{Cu}_2\mathrm{OSeO}_3$ \cite{SekiScience2012}), leading to the so-called bulk DMI. Broken inversion symmetry can also exist at the interfaces of magnetic thin films and multilayers \cite{RonaldNatReviewMater2016}, leading to the interfacial DMI where the coupling between two atomic spins ($\mathbf{S}_1$ and $\mathbf{S}_2$) in a thin magnetic layer is mediated by a third atom with strong SOC in a neighboring non-magnetic or magnetic layer. This opens up potential for tailoring the interfacial DMI by materials design (heterostructure, strain, chemical doping), which has been demonstrated in ferromagnet/heavy-metal \cite{RonaldNatReviewMater2016}, antiferromagnet/heavy-metal \cite{AkandaPRB2020,JaniNatureFe2O3Pt2021,YuhanLingAdvSci2023}, and recently, insulating magnets based system (Pt/TmIG/GGG \cite{DingPRB2019TmIGGGG} and substrate/rare-earth(RE)IG/metal trilayer \cite{CarettaNatComm2020}), and synthetic antiferromagnets \cite{Cowburn_InterlayerDMI_NatMater2019,LegrandFertNatMater2020} where two atomic spins from two neighboring ferromagnetic layers can even have DMI through the large-SOC atoms in the paramagnetic spacers \cite{Cowburn_InterlayerDMI_NatMater2019}. Such interlayer-type interfacial DMI is thus different from the conventional intralayer-type interfacial DMI where the two atomic spins reside in the same magnetic layer.

The manifestation of DMI in magnon-magnon coupled system has remained scarce and been limited to synthetic antiferromagnet (syn-AFM) based systems \cite{YuanPRB2023,wangPRB2024ultrastrongtheory}, where the ultrastrong magnon-magnon coupling is predicted by tailoring the interplay among the magnetic anisotropy field, the effective interlayer DMI field, and the effective RKKY interaction field \cite{wangPRB2024ultrastrongtheory}. At present, it remains unclear how the synergistic interaction among DMI, direct exchange interaction, and RKKY interaction at the interface of two magnetically ordered layers influences the magnon-magnon coupling. To address this fundamental question, it is imperative to develop low-damping magnetic heterostructures with tunable spin-spin interactions of different types (DMI, direct exchange, and RKKY). One approach is to insert a composite interface layer \cite{cuchet2016perpendicular} such as Fe$_{1-x}$Pt$_x$ between the Py and YIG. 

\textit{A multi-coupling scheme} Magnon-magnon coupling in the presence of a combined, synergistically-controlled interfacial coupling scheme is a particularly interesting material engineering direction for hybrid magnonics. \textcolor{black}{For example, combining analytical theory and magneto-optical measurement, a recent work \cite{christy2025tuning} has demonstrated the capability of exploiting the synergistic effect of Zeeman torque and exchange torque at a Py/YIG interface to tune the spin precessional phase of the PSSW mode in the YIG, hence opening new potential application as phase-controlled hybrid magnonics. Looking ahead, using the above-mentioned Py/Fe$_{1-x}$Pt$_x$/YIG heterostructure as an example, \textit{how to understand, predict, and characterize the synergistic effect of interfacial DMI and direct exchange coupling on the phase, amplitude, and wavevector-dependence of the magnons in the YIG?}} 

\textcolor{black}{From a theoretical/computational perspective, addressing this question requires a new analytical or micromagnetic model that can predict the magnetic susceptibility and coupled magnon-magnon dynamics in a multiphase system with  coexistence of interfacial exchange coupling and interfacial DMI at the interface of two dissimilar magnetic layers, and can be further coupled to photon and/or acoustic phonon subsystems. To this end, the first step would be to develop a generalized magnetic boundary condition that simultaneously allows a partial transfer of exchange torque, the magnetic energy flow conservation across the interface, and the minimization of the magnetic interface energy (which is contributed by both the isotropic exchange energy and the anisotropic DMI energy\cite{hu2018strainskyrmion}).} 

\textcolor{black}{To date, existing theories and computational models only permit a rigorous treatment of the boundary conditions associated with the direct exchange coupling between two magnetic layers (e.g., the Hoffman boundary condition \cite{ZZhangPRB2021,LiuPRB2024TmIG/YIG}), the interfacial DMI between a magnetic layer and a non-magnetic layer \cite{RohartPRB2013,SampaioNatNanotechnol2013}, and more recently, the interlayer DMI in a Syn-AFM \cite{ElenaPhysRevAppl2025}. Although a generalized Barnas-Mills boundary condition has been developed for a finite-thickness interface with the coexistence of DMI and direct exchange interaction \cite{KruglyakJPCM2014}, such a boundary condition has not yet been implemented in any analytical and numerical studies related to magnon-magnon coupling or evaluated in comprehension}.

\section*{Future perspective of magnon-magnon hybrid materials system}

\begin{figure*}[htb]
\centering
\includegraphics[width=6.8 in]{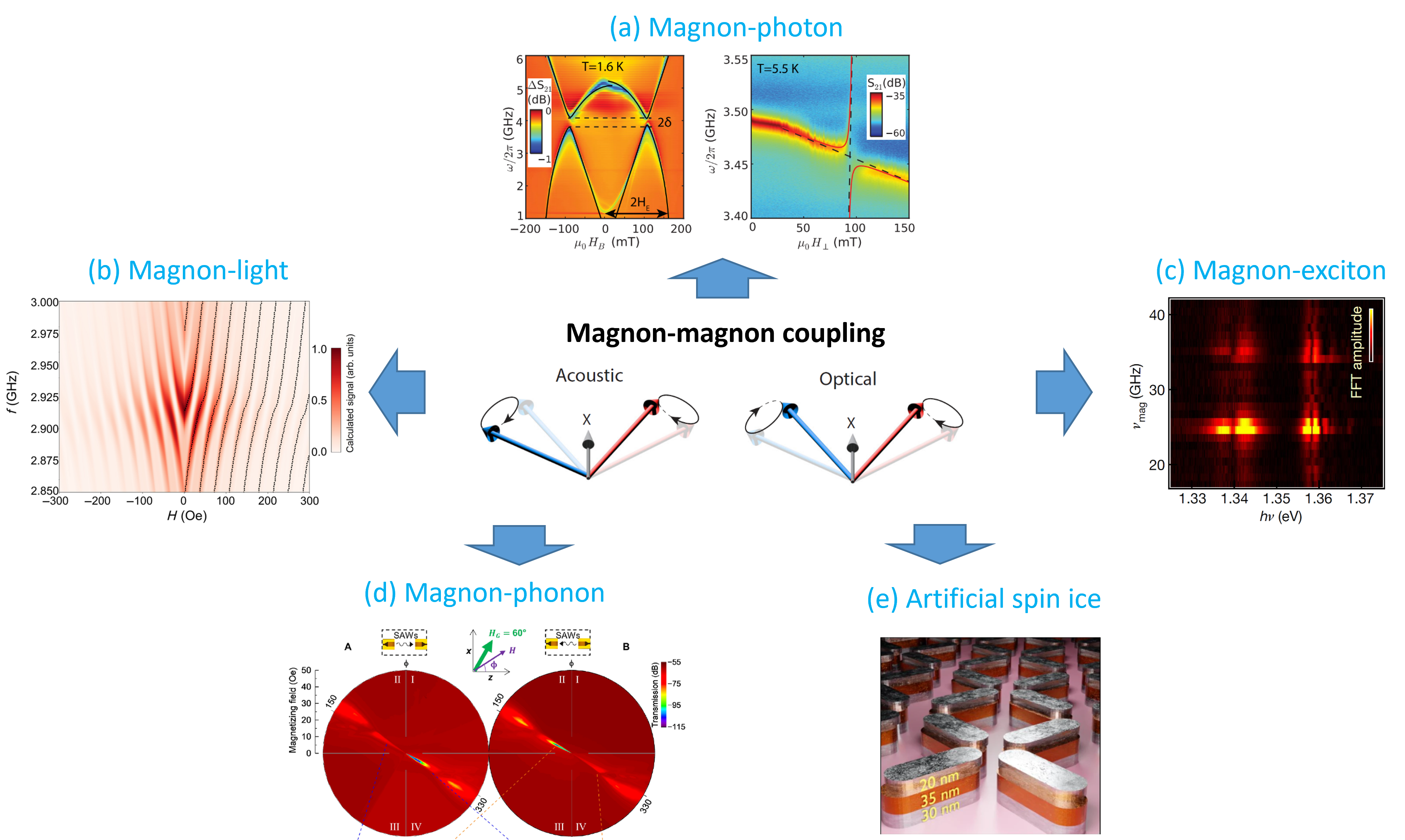}
\caption{Future perspective of magnon-magnon coupling as an approach to modulate interactions in other hybrid magnonic systems. Examples including (top to bottom): (a) tuning magnon-photon interaction via magnon-magnon gap, figure taken from Ref. \cite{LiPRR23}; (b) modulating magnon-light interaction and spectrum using magneto-optical effect and its inverse, figure taken from Ref. \cite{inman2022hybrid}; (c) synergy with magnon-exciton modulation, figure taken from Ref. \cite{bae2022exciton}; (d) nonreciprocal magnon-phonon interaction due to magnon-magnon coupling, figure taken from Ref. \cite{MPageSciAdv2020}; (e) artificial spin ice system leveraging magnon-magnon coupling, figure taken from Ref. \cite{dion2024ultrastrong}.}  
\label{fig_persp}
\end{figure*}

The discovery of magnon-magnon coupling in various material systems offers a new approach to modulating and controlling intrinsic magnon excitations, potentially enabling new applications in coherent magnon engineering across a wide range of hybrid magnonic platforms. Below we highlight several representative hybrid magnonic systems that incorporate magnon-magnon hybrid materials, and explore how magnon-magnon coupling can enhance and expand the capabilities of magnon-based coherent information processing, summarized in Fig. \ref{fig_persp}.

\subsection*{Magnon-photon interaction}

\textit{Strong coupling and magnon-magnon gap} Magnon-photon coupling is the fundamental interaction in cavity magnonics, originating from the strong magnetic dipolar coupling between magnetic excitations and electromagnetic waves. Through cavity-enhanced interaction, strong magnon-photon coupling can be achieved, enabling coherent energy transduction between magnons and microwave photons \cite{SoykalPRL10,CaoPRB15,ZARERAMESHTI20221}. This mechanism forms the foundation of quantum magnonics \cite{lachance2019hybrid}, where magnon-qubit entanglement can be realized, mediated by cavity microwave photon acting as a coherent data bus. 

By incorporating magnetic materials with magnon-magnon coupling, one can modulate the efficiency of magnon-photon coupling via the magnon-magnon bandgap. 
An example is the layered hybrid perovskite antiferromagnet (CH$_3$CH$_2$NH$_3$)$_2$CuCl$_4$ (Cu-EA) \cite{comstock2023hybrid}. In Cu-EA, \textcolor{black}{the intrinsic interlayer coupling features antiferromagnetic type. In addition, the structure-induced DMI breaks the symmetry, and hence leads to the hybridization of interlayer acoustic and optical magnon modes, manifesting a magnon-magnon bandgap at their crossing point.} In addition, the position of this magnon bandgap can be sensitively tuned by temperature as a result of changes in the interlayer antiferromagnetic coupling strength. When the frequency of a resonator mode lies within the magnon–magnon bandgap, the magnon excitations become decoupled from the microwave photons, effectively disabling the magnon-photon interaction. This effect was demonstrated by Li \textit{et al}. \cite{LiPRR23}, where a Cu-EA crystal was coupled to a coplanar superconducting resonator in a flip-chip configuration. As the temperature varies, the magnon–photon anticrossing gap closed and reopened as the magnon-magnon band gap shifted across the resonator frequency. Consequently, the magnon–photon coupling strength varied from 35 MHz at 1.5 K, dropped to zero between 3 and 3.5 K, and then increased to 42 MHz at 6 K. In particular, the temperature range of zero coupling can be engineered by designing the resonator frequency to coincide with the desired location within the magnon–magnon bandgap.

The demonstration highlights the opportunity to manipulate coherent mode hybridization and control coherent information processing with quantum properties in complex magnetic materials. Another similar example is to use a superconducting resonator to investigate magnon-magnon coupling between two remote quasi-two-dimensional topological magnets Cu[1,3-benzenedicarboxylate(bdc)] via their mutual magnon-photon coupling to the resonator \cite{PengLiAPL25}. In addition, the use of cavity magnonics also provides a new way to probe complex \textcolor{black}{spin-wave} dispersion with narrow-band microwave characterizations instead of broad-band ferromagnetic resonance. Similar ideas can also be implemented with other magnon-magnon hybrid materials with different magnon bandgap engineering.

\textit{Purcell regime and superstrong coupling} Another example is when the magnon-magnon coupling is in the intermediate regime, in which one magnon mode simply enhances or attenuates the excitation of the other magnon counterpart, e.g. the cavity modes in the confined geometry. In this case, as introduced earlier, a magnonic cavity can be formed \cite{inman2022hybrid}. The excited cavity modes can then interact with additional photon mode in a multi-partite coupling scheme. In the earlier example by Inman \textit{et al}., a photon resonator mode is provided by a split-ring resonator, which couples to a magnon-magnon coupled YIG/Py bilayer. The Kittel mode of YIG couples to the photon mode directly, forming a strong coupling anti-crossing. In addition, the cavity modes of YIG (prompted by the Py) can also couple to the photon mode, exhibiting a superstrong coupling spectrum, \textcolor{black}{i.e. when the coupling strength is of the same order of the free-spectral range (FSR) \cite{meiser2006superstrong}, and in the present example, the FSR can be neatly tuned by the cavity size (thickness of YIG).}

\subsection*{Magnon-light interaction}



Optical light, known for its low-loss propagation, resilience to environmental noise and high bandwidth, is the backbone of modern communication systems and shows strong potential for the next-generation information processing and transduction such as photonic integrated circuits \cite{ShekharNC24}, quantum optics \cite{MaringNPho24} and quantum communication \cite{AwschalomPRXQ21}. While electro-optic \cite{PanicciaNature04} and opto-mechanical \cite{AspelmeyerRMP14} systems are the main-stream approaches for converting optical information to solid-state signals, magneto-optical systems have emerged as promising alternatives due to their broad frequency tunability and unique nonreciprocity for implementing on-chip photonic isolators \cite{BiNPho11,ZhangOptica19,Srinivasan22}. Magnon-light interactions can occur through magneto-optical coupling such as the Faraday effects in bulk materials and Kerr effect at the interface, where the light polarization is modified by the magnetic moment and thus modulated by the magnetic excitations \cite{HisatomiPRB16,Kaffash2023}. This process has been demonstrated using YIG sphere resonators via interactions between the Kittel magnon mode and whispering-gallery optical modes \cite{OsadaPRL16,ZhangPRL16}. Recent advances target on-chip optomagnonic resonators based on single-crystalline YIG \textcolor{black}{thin films \cite{zhu2020waveguide,wu2025microwave}}, which offer enhanced mode overlap and reduced volume, enabling stronger and more efficient magnon-optical interactions.

The introduction of magnon-magnon hybrid material system can highly enrich magnon-light modulation. By depositing a metallic ferromagnetic thin film, such as Py, on one side of a YIG film, the Py layer acts as a mirror which can reflect optical light that is normal incident from the other side of the YIG film. This YIG/Py bilayer acts as a magneto-optical modulator where both the Faraday effect in YIG and the Kerr effect at YIG/Py interface contribute to the light polarization rotation \cite{xiong2020probing,xiong2022tunable,christy2025tuning,zhou2025magneto,xiong2024phase}. Note that due to the sign reversal of both the light chirality and wavevector during reflection, the Faraday effect of the reflected light in YIG contributes with the same sign as the incident light, doubling the Faraday signal. The coherent coupling between YIG and Py changes the symmetry of the YIG PSSW modes with short wavelengths, allowing them to couple with external uniform microwave field. This has led to a series of new coherent phenomena in magneto-optical interactions, such as \textcolor{black}{magnetically-induced transparency \cite{zhang2014strongly,xiong2020probing,xiong2022tunable} and zero-reflection \cite{qian2023non,christy2025tuning}}. Specifically, the magnon-magnon coupling can:

(i) allow to leverage the various interfacial coupling types with distinct symmetries (direct-exchange, RKKY, and DMI) to operate in a single or collective fashion, so as to fully exploit the rich spin wave dispersions (beyond just Kittel magnon mode). This can be realized via engineering the film(or nanostructure)'s size and dimension, excitation geometry, and magnetic field. 

(ii) make the overall concept of using film-based structures more compatible with the state-of-the-art photonic and phononic architectures, and can thus be seamlessly integrated in an all-on-chip format. In particular, the magnonic waveguide resonators can be further subject to additional electrical, optical, or mechanical control knobs (the available CMOS toolkit) \textcolor{black}{\cite{wang2024nanoscale,merbouche2024true,au2023electric,qin2021nanoscale,breitbach2023stimulated}}. 

(iii) provide a platform for accessing and controlling magnon-light transduction exploiting unconventional magnon modes or collective states, such as the magnon BEC mode, in the presence of strong nonlinear magnon processes, such as three- and four-magnon scatterings \cite{qu2025pump}.  

\subsection*{Magnon-phonon interaction}

The coupling between magnons and acoustic phonons represents the dynamical energy exchange between the spin and lattice subsystem of a material. Similarly to other hybrid magnonic systems, the coupling strength represents the rate of the energy exchange, which is proportional to the magnetoelastic coupling constant and the mode profile overlap \cite{LiAPLMater2021,HuMRSBulletin2024,Lee1955magnetostriction}. Since magnetoelastic coupling coefficient is a \textcolor{black}{fourth-order} tensor, this interaction energy density is essentially nonzero in any materials that have a nonzero net magnetization in the entire body (ferro-/ferri-magnets) or its sublattice (ferri-/antiferro-magnets). This universality provides flexibility in the selection of materials platform for studying magnetoelastic interactions. Additionally, the frequency and wavenumber of the magnons and acoustic phonons for most magnetically ordered materials are not too distinct in the GHz-THz range, making it efficient to either excite magnons by acoustic phonons \textcolor{black}{\cite{vlasov2022modern,ImagingMagnetoacoustic2020,BombeckPRB2012GaMnAs,ThevenardPRB2010GaMnAsP,GhitaPRB2023Anatomy,mocioi2023towards,VernikPRB2022,KAnPRB2010YIG,JingXuPRAppl2021,zhuangACSApplMaterInterfaces2021,zhuangnpjComputMater2022,zhuang2023acoustic,ZhuangPRApp2024}} and vice versa. For example, by leveraging the developments from the field of microelectromechanic systems, such as the surface acoustic wave (SAW) or bulk acoustic wave (BAW) based devices, and the field of picosecond ultrasonics \cite{Matsuda2015Review}, efficient acoustic excitation of magnons has been demonstrated experimentally at both the GHz and THz regime. Due to such flexibility in materials and device design, hybrid magnon-phonon system has emerged as one of the most intensively studied hybrid magnonic systems, which has been utilized to realize coherent gate operation for quantum information science applications \cite{LiJAPReview2020,JingXuPRAppl2021}, enable long-distance (mm-scale) magnon transport \cite{ImagingMagnetoacoustic2020}, and design dynamically tunable, narrowband THz emitters \cite{HuMRSBulletin2024,zhuangnpjComputMater2022,zhuangACSApplMaterInterfaces2021} and THz optoelectronic transducer \cite{ZhuangPRApp2024} (see detailed discussion of device architectures in relevant patents \cite{US11112355B2,US11199447B1,US11817242B2,US12345930B2}). 

The combination of magnon-magnon coupling with magnon-phonon coupling has enabled emergent functionalities, such as the nonreciprocal transmission of SAWs in a hybrid system comprised of a FeGaB/Al$_2$O$_3$/FeGaB multilayer integrated onto a LiNbO$_3$-based SAW device \cite{MPageSciAdv2020}. Specifically, the magnetic dipolar coupling between the two FeGaB layers gives rise to asymmetric magnon dispersion relation with respect to positive and negative wavevectors. As a result, for a fixed frequency, the resonant magnon-phonon interaction, where the frequencies and wavenumbers of both the magnons and acoustic phonons are identical, only occurs when the wavevector is, for example, negative. This would then lead to strong absorption of an SAW (i.e., its energy is transferred to the magnetic subsystem) that has a negative wavevector, whereas the SAW with a positive wavevector can transmit with much lower level of absorption. In a similar spirit, we expect that the magnon-magnon coupling, especially the synergistic interaction of different types of spin-spin coupling (direct exchange, DMI, RKKY, etc.), can further enrich the magnon-phonon coupling by introducing new features in the magnon dispersion relation and hence its overlap with the acoustic phonon dispersion. \textcolor{black}{We suggest that} recently developed TmIG/YIG multilayer \cite{TmIGYIGPRAppl2024} and rare-earth iron garnet superlattice such as TbIG/TmIG \cite{RossTbIG/TmIGSLs2024} \textcolor{black}{are promising material candidates for studying such interplay between magnon-magnon coupling and magnon-phonon coupling for three reasons. First, the possible coexistence of direct exchange and DMI at the garnet/garnet interface enables the study of such a competing spin-spin interaction on magnon-phonon coupling. Second, epitaxial strain can be used to tune the orientation of ground-state magnetization in such iron garnet thin films, such as the introduction of perpendicular magnetic anisotropy\cite{RossTbIG/TmIGSLs2024,das2023perpendicular,DingMingzhongPRAppl2020} or noncollinear magnetization configuration, thereby enabling new functionalities into both magnon-magnon coupling (see \cite{FanRossPRL2025} for a recent example) and magnon-phonon coupling. Third, it is possible to simultaneously engineer the magnon and phonon dispersion relation by superlattice design via band folding}.

\subsection*{Magnon-exciton interaction}

An exciton is a bound state of a pair of electron and hole which are coupled via direct Coulomb interaction in a solid. In two-dimensional semiconductors, many show strong exciton excitations due to the significantly reduced dielectric screening of Coulomb interaction. These excitions couple strongly to light and persist to room temperature, allowing rich features in physics to be explored optically. In addition to exciton-light coupling, excitons can also couple to magnons due to spin-dependent electron and hole transfer, with the recent example of 2D magnetic semiconductor CrSBr \cite{bae2022exciton}. This novel optical approach of probing magnon-electron interaction has been widely utilized to investigate polariton-magnon coupling \cite{dirnberger2023magneto}, \textcolor{black}{propagating magnons \cite{bae2022exciton,sun2024dipolar}}, strain effects \cite{cenker2022reversible} and magnetic-field-controlled magnon-magnon hybridization \cite{DiederichNNano23}, as well as higher-harmonic generation \cite{DiederichNNano25}. The degree of magnon-exciton hopping is determined by the spatial and spinor components of the exciton wavefunction overlap between layers. Since the structure is fixed, the spatial wavefunction overlap is fixed and the spinor part can be tuned using an external magnetic field \cite{wilson2021interlayer}. Among various 2D magnetic semiconductors, this coupling is only observed in CrSBr because of its highly dispersive electronic band structure, leading to Wannier-type, delocalized excitons \cite{bae2022exciton}. For instance, the magnetic semiconductor NiPS$_3$ and the magnetic insulator CrI$_3$ exhibit spatially localized, Frenkel-type excitons, and do not possess the similar spin-dependent charge transfer\cite{wu2019physical,dirnberger2022spin,brennan2024important}.

The introduction of magnon-magnon coupling in CrSBr, due to its nature of layered van der Waals antiferromagnet \cite{cham2022anisotropic}, brings a new approach to modulate magnon-exciton coupling by controlling the symmetry of the magnon-magnon hybrid modes. Recently, Diederich \textit{et al}. \cite{DiederichNNano23} demonstrated that coexisting magnon-magnon and magnon-exciton couplings can turn a dark acoustic magnon mode into a bright mode. Without magnon-magnon coupling, only the optical mode is bright because the optical mode involves changes in canting angle, whereas the acoustic mode has a fixed canting angle. However, applying an external field breaks the two-fold rotational symmetry and creates a hybridized mode between acoustic and optical magnons \cite{cham2022anisotropic}. When the two modes are hybridized, both become bright and can be easily detected using the exciton sensing method. The coexisting magnon-magnon and magnon-exciton couplings in CrSBr also allow efficient generation and detection of nonlinear magnon dynamics such as magnon high-harmonic generation, parametric amplification, and sum- and difference-frequency generation \cite{DiederichNNano25}.

Besides magnon-exciton coupling, another feature of CrSBr is its strong magnetoelastic coupling \cite{BaePRB24}. Compared to three-dimensional magnets, two-dimensional magnetic systems often display enhanced magnetoelastic interactions due to the confinement of spin interactions within the in-plane direction \cite{GuNL22}. How these multiple coexisting couplings interact to give rise to new magnonic phenomena remains an open and intriguing area of research.

\subsection*{\textcolor{black}{Floquet engineering of hybrid magnonic systems}}

\textcolor{black}{Floquet engineering, the application of a time-periodic external field, has become an efficient method to control electronic and spintronic properties and create new coherent states in quantum materials \cite{Oka2018FloquetReview}. In hybrid magnonic systems, Floquet engineering has thus far been implemented via the application of either a continuous sinusoidal magnetic field or periodic pulses. Several new functionalities such as Floquet-induced magnonic Autler-Townes splitting \cite{XufengPRL2020Floquet} and on-demand dark-bright mode conversion \cite{XufengPRApp2025BrightDarkmode} have been experimentally demonstrated in hybrid magnon-photon systems based on bulk YIG spheres. Furthermore, Floquet control of magnonoic Rabi oscillation (where the Rabi flopping frequency between two magnon polariton modes was tuned by varying the amplitude of continuous Floquet drive) and magnonic Ramsey interference (where the relative phase difference between the two modes was controlled by frequency detuning and the control of free evolution time under pulsed Floquet drive) have been computationally demonstrated in practical-sized 3D hybrid magnon-phonon cavity based on YIG spheres \cite{zhuang2024dynamical}. Looking ahead, driven by the need of miniaturization, power reduction, and on-chip heterogeneous integration, we envisage opportunities and challenges rising in (i) the extension of Floquet Engineering to on-chip hybrid magnonic systems (especially the magnon-phonon cavity), and (ii) the adoption of Floquet fields that are easier to localize on a chip such as r.f. voltage-induced surface acoustic waves and current-induced spin torques.}

\textcolor{black}{Beyond device-level control, Floquet engineering enables access to intrinsically nonequilibrium phenomena with no equilibrium analogue. One prominent example is the creation of a synthetic frequency (time) dimension, in which harmonics of a driven mode constitute a lattice and the Berry curvature defined over momentum–frequency (Sambe) space governs topological energy transfer between tones—a generalized Thouless pump in frequency space \cite{martin2017topological}. Although explored in photonics and other quantum platforms, this framework carries over naturally to cavity magnonics, where strong coherent magnon–photon coupling and periodic control of coupling and detuning have been demonstrated \cite{xu2020floquet}, and the interplay between coherent and dissipative couplings under Floquet driving  has been analyzed in detail theoretically \cite{yang2023theory}. 
A complementary opportunity is offered by leveraging the interplay of periodic drive and dissipation to realize non-Hermitian Floquet phases, including PT‑symmetric and PT‑broken regimes characterized by exceptional points, mode coalescence, non‑orthogonal eigenmodes, and enhanced susceptibility \cite{liu2024floquet}. This perspective has recently been translated to microwave cavity platforms, where tailoring the time dependence of dissipation in tandem with Floquet modulation implements a non‑Hermitian shortcut to adiabaticity, enabling fast, high‑fidelity state transfer \cite{zhang2022non}. Although this exploration is only beginning—particularly on the experimental front—the convergence of Floquet control, hybrid magnonics, and engineered dissipation clearly signals substantial potential for realizing and harnessing nonequilibrium phases and functionalities beyond equilibrium platforms.}

\subsection*{Computing with artificial spin ice}

Artificial spin ice (ASI) \textcolor{black}{\cite{Sultana_2025,gliga2020dynamics, sklenar2019dynamics, skjaervo2020advances, wang2006}} systems are increasingly recognized as promising physical platforms for neuromorphic computing owing to their vast microstate landscapes, intrinsic non-linear dynamics, and reconfigurability \cite{saccone2019towards,SkjarvoNRP20,LendinezJCPM20,GligaAPLMater20,GartsideNNano22,Vanstone_2022,bhat2023spin,StenningNC24}. \textcolor{black}{ASI  consists of lithographically fabricated 2D arrays of nanomagnets arranged in periodic or aperiodic lattices, offering a versatile platform to study magnetic states absent in natural materials. Initially conceived as mesoscopic analogs of frustrated pyrochlores \cite{harris1997geometrical, bramwell2001spin}, ASI has since evolved into a rich field of its own.}

\textcolor{black}{Typically, each nanomagnet acts like a binary Ising-like macrospin by selecting the aspect ratio of individual nanomagnets in these lattice to ensure a single-domain state \cite{Sultana_2025}. However, for certain applications, a bistable macrospin–vortex configuration in individual nanomagnets can be advantageous \cite{GartsideNNano22,dion2024ultrastrong}.} The collective behavior at lattice vertices enables direct access to frustration, emergent monopole–antimonopole excitations, and phase transitions. These systems also hold promise as reprogrammable magnonic crystals, where spin waves function as coherent information carriers \cite{rana2021applications, lendinez2019magnetization, gubbiotti20242025, barman20212021}.

\textcolor{black}{Moreover, t}he collective interactions within these nanomagnet arrays naturally support paradigms like reservoir computing, processing temporal information efficiently by harnessing internal dynamics \cite{GartsideNNano22,KaffashPLA21}. Extending these systems into three dimensions (3D), especially via vertically stacked layers, significantly enhances their potential \cite{LadakAPLMater22,PachecoNC17,dion2024ultrastrong}. Recent work demonstrates that purely dipolar coupling between layers can achieve ultrastrong magnon-magnon interaction without direct exchange, enabling highly tunable magnonic metamaterials that overcome traditional coupling-reconfigurability trade-offs \cite{dion2024ultrastrong,NoriNRP19}. This 3D approach allows for substantial zero-field mode shifts, the generation of distinct hybrid modes (acoustic/optical), and phase control, offering unprecedented spectral manipulation for future devices and enhanced neuromorphic capabilities through richer dynamics and exponentially larger state spaces (e.g., $16^N$ vs $2^N$) \cite{GartsideNNano22,Vanstone_2022,StenningNC24,dion2024ultrastrong}.

Looking ahead, the refined control offered by 3D ASI opens avenues beyond conventional neuromorphic computing into areas like spin logic and wave-based processing \cite{PirroNRP21}. The ability to program distinct microstates, each with a unique magnonic spectral fingerprint, improves readout and computational power \cite{Vanstone_2022,gartsideNatureCommunications2021}. Furthermore, functionalities like chirality-selective control of vortex states using inter-layer dipolar fields act as programmable switches, introducing spectral asymmetries useful for low-field sensing or information encoding \cite{dion2024ultrastrong}. Manipulating hybridized modes with specific phase relationships is also key for coherent magnonics, potentially enabling phase-based logic and directional information flow \cite{PirroNRP21}. Future research will likely focus on integrating more layers or diverse materials (including antiferromagnets or active spacers, such as Ru, and 2D magnets) and further exploring these chiral and non-reciprocal effects within complex 3D structures to unlock new functionalities \cite{MayNC21,SacconeCP23}.

\section*{Conclusion}

As a new subfield of hybrid magnonics, magnon-magnon coupling focuses on the fundamental interactions between magnon modes through material engineering. This includes interfacial exchange engineering in magnetic bilayers, magnon band structure engineering in van der Waals antiferromagnets or synthetic antiferromagnets, and dipolar field engineering in nanomagnet arrays. These magnetic interactions offer a means to control and design magnon eigenmodes, akin to electron band structure engineering but accessible via electrical excitation and detection at microwave frequencies. Moreover, since magnon-magnon coupling occurs within engineered magnetic materials, it can be seamlessly integrated with other hybrid magnonic systems, enabling the potential modulation of magnon-$X$ coupling through magnon band engineering. 

\textcolor{black}{As a summary of the section of future perspectives, we anticipate three major directions for future developments in magnon-magnon coupling, as i) creating new functionalities for magnon-based coherent information processing and transduction in coupling to other dynamic systems, such as microwave photons, acoustic phonons, optical light, and excitons; ii) investigating fundamental phyics in novel quantum materials, such as spin--electron (magnon-exciton) coupling in layered van der Waals antiferromagnets, and Dzyaloshinskii--Moriya interaction in hybrid organic perovskite antiferromagnets; iii) advancing spin-wave computation schemes using artificial nanomagnetic networks.} In all three directions, we remark on the importance of identifying suitable materials or hybrid material systems with large magnon-magnon coupling strengths and low magnon dampings for achieving strong coupling or ultrastrong coupling. Promising examples of low-damping materials include YIG and TmIG (magnetic insulators), Py and CoFeB (metallic ferromagnets), CrCl$_3$ and CrSBr (layered van der Waals antiferromagnets). \textcolor{black}{In dipolar- and exchange-governed systems, additional} structural engineering, such as reducing the film thickness or decreasing the spacing between adjacent nanomagnets, can also boost the coupling strength. As a conclusion, magnon-magnon coupling is a rapidly growing field expanding in materials science, quantum engineering, and with high potential of developing into a dynamic interdisciplinary field in hybrid magnonics.


\section*{Acknowledgments} 

W.Z. and Y.X. acknowledge support from the National Science Foundation under Grant No. NSF DMR-2509513. J.-M.Hu acknowledges support from the National Science Foundation under Grant No. DMR-2237884 and partial support for manuscript preparation from the Wisconsin MRSEC (DMR-2309000). M.M.S and J.S. acknowledge support from the National Science Foundation under DMR-2328787. M.B.J. acknowledges support from the National Science Foundation under Grant No. 2339475. V.S.B. was supported by the U.S. Department of Energy, Office of Science, Office of Basic Energy Sciences under Award Number DE-SC-0024346. L.L. and Q.W. acknowledge support from National Science Foundation under Grant No. ECCS-2309838. The effort from Y.L. in paper writing was support by the U.S. DOE, Office of Science, Basic Energy Sciences, Materials Sciences and Engineering Division under contract No. DE-SC0022060. B. F. acknowledges support from the National Science Foundation under Grant No. NSF DMR-2144086. Y.K.L. acknowledges support from U.S. Department of Energy, Office of Basic Energy Sciences (DE-SC0025422).  

\section*{Competing Interests}

W.Z., Y.X., J.-M.H., J.S., M.M.S., V.S.B., Y.L., L.L., Q.W., Y.K.L., Y.J.B., B.F. declare no financial or non-financial competing interests. M.B.J. serves as an Editor of this journal and had no role in the peer-review or decision to publish this manuscript. M.B.J. declares no financial competing interests.

\section*{Author Constributions}

W.Z., Y.L., J.H., J.S., B.J. conceived the project and wrote the initial version of the manuscript. W.Z., Y.L. organized the figures and the table. All authors contributed to the writing and editing of the manuscript. All authors have read and approved the manuscript.

\section*{REFERENCES}  
\bibliography{sample}

\section*{Figure Legends}

\textbf{Figure 1.} The field of hybrid magnonics has grown and developed into different ramifications, such as magnon-photon, magnon-phonon, and magnon-light coupling systems, with the goal of developing energy and signal transduction functionalities across different physical platforms. The magnon-magnon coupling is presented as a unique and versatile approach towards inducing and tailoring magnon modes with desirable properties for further hybridization in the various hybrid magnonic contexts, i.e. Magnon + $X$. \textcolor{black}{In an analogous Bloch sphere representation, the ``magnonic Ramsey process'' incurs two $\pi/2$-pulses that prompt a non-interaction zone and open a free-precession regime between the two-level states, enabling the magnon phase as a state control variable, in contrast to the Rabi process, which only shuffles states directly between the two interaction zones.} 

\textbf{Figure 2.} The various key attributes of magnon-magnon coupling in the context of hybrid magnonics. 

\textbf{Figure 3.}  (a) Schematic of coupled magnon-magnon dynamics in a YIG/FM bilayer mediated by interfacial exchange coupling, with uniform mode ($k=0$) excited in the FM layer and the perpendicular standing spin wave mode ($k>0$) excited in the YIG layer. (b) $H_r$-$\omega$ dependence for the YIG PSSW mode (red) and the FM uniform mode (blue) that intersect with each other. YIG uniform mode is also shown in red dashed curve. (c-f) Coherent magnon-magnon coupling in (c) YIG/Co \cite{klingler2018spin}, (d) YIG/Ni\cite{chen2018strong}, (e) YIG/CoFeB\cite{qin2018exchange} and (f) YIG/NiFe\cite{li2020coherent}.

\textbf{Figure 4.} The various geometries of magnon-magnon coupling in bilayer magnetic heterostructures. (a,b,c): uniform, extended bilayer coupling in which the thickness of the film defines the magnonic cavity, for (b) insulator/metal (Figure taken from Ref.\cite{xiong2020probing}), and (c) all-insulator systems (Figure taken from Ref. \cite{liu2024strong}). (d,e,f): the device geometry in which an FM stripe couples locally to an extended YIG film. The excited magnons can propagate along the lateral dimension, and be detected via: (e) dc rectifications, e.g., using spin Hall effect of Pt (Figure taken from Ref. \cite{sheng2023nonlocal}), or (f) with rf inductive antennas and magneto-optical probes (Figure taken from Ref. \cite{qin2021nanoscale}). (g,h,i): lateral magnonic cavity in which the FM stripes (h) serve as complementary, local antennas (Figure taken from Ref. \cite{liu2018long}), or (i) only used to define the cavity boundaries while the excitation is globally sourced otherwise (Figure taken from Ref. \cite{santos2023magnon}). 

\textbf{Figure 5.}  Detecting magnon-magnon coupling in 2D AFM systems. (a) Schematic of microwave absorption experiment with 2D material. (b) Right-handed (RH) and left-handed (LH) magnon modes. (c) Optical and acoustic magnon modes. (d) Strong magnon-magnon coupling between RH and LH modes in CrSBr. (e) Strong magnon-magnon coupling between optical and acoustic modes in CrCl$_3$. The coupling is only activated by breaking in-plane rotational symmetry. (f) Magnetic state alters the exciton resonance energy in CrSBr. (g) Bright and dark magnon modes. (h) Magnon dispersion detected with optical reflectivity measurements in CrSBr. (i) Calculated transient exciton resonance energy shift from the optical (red) and acoustic (blue) magnon modes. (j) Coherent magnon hybridization between bright and dark magnon modes detected by time-resolved measurement. (Figures (a-d) taken from Ref. \cite{cham2022anisotropic}; (e) taken from Ref. \cite{macneill2019gigahertz}; (f-j) taken from Ref. \cite{diederich2023tunable}.)

\textbf{Figure 6.} Detecting magnon-magnon coupling in bulk AFM systems. (a) Schematic of time-resolved THz free induction decay (FID) measurement. (b) Sample geometry of YFeO$_3$ in a tilted magnetic field to study the THz frequency magnon bands. (c) Calculated co-rotating (blue) and counter-rotating (red) coupling strengths show the dominance of the counter-rotating coupling terms. (d) Measured THz magnon bands and the ultra-strong magnon-magnon coupling in YFeO$_3$. (e) quasi-FM (qFM) and quasi-AFM (qAFM) magnon mode. (f) Excitation of different magnon modes in ErFeO$_3$. (g) Strong magnon frequency up-conversion illustrated in 2D THz FID spectra. (h) Dependence of the magnon up-conversion signal amplitude on the pump magnetic field. (i) Magnon energy-level diagrams. (j) The 2D THz spectra in YFeO$_3$ summarizing the rich linear and nonlinear processes including pump-probe (PP),  rephasing or photon echo (R), non-rephasing (NR), two-quantum (2Q), second harmonic generation (SHG), sum-frequency generation (SFG) and difference frequency generation (DFG). (Figures (a),(e),(i),(j) taken from Ref. \cite{zhang2024terahertznonlinear}; (b-d) taken from Ref. \cite{makihara2021ultrastrong}; (f-h) taken from Ref. \cite{zhang2024terahertzupconversion}.)

\textbf{Figure 7.}  Interactions between acoustic and optical magnons in synthetic magnets are summarized above. The upper left portion of the figure illustrates how, for synthetic antiferromagnets, the crossing between an acoustic and optical branch [(a)] can turn into an avoided crossing by changing the wave number [(b)], or applying a symmetry breaking external field oblique to the sample plane, [(c), (d)].  The respective references for these images are taken from \cite{sud2023magnon}, \cite{shiota2020tunable}, \cite{sud2020tunable}, and \cite{dai2021strong}. Similar avoided crossings are shown in the lower left portion of the figure in a synthetic ferrimagnet, where the magnetic layers have unequal thicknesses or dissimilar magnetizations, [(e), (f)] taken from Ref. \cite{hossain2024broken,sud2023magnon}. An alternative strateegy to engineer interactions into these materials is by changing the number of magnetic layers. The right portion of the figure, [(g), (h)] taken from Ref. \cite{rong2024layer,subedi2025engineering}, shows how in a tetralayer, acoustic or optical magnon pairs interact. If interlayer spin pumping is present in the tetralayer, optical and acoustic magnon pairs can interact as well without the aid of a symmetry breaking field.

\textbf{Figure 8.}  (a), (b) Phase-resolved TR-MOKE microscopy maps and line profiles measured on a 100-nm-thick Y\textsubscript{3}Fe\textsubscript{5}O\textsubscript{12} (YIG) film with a 730-nm-wide CoFeB stripe. Figure taken from ref. \cite{Qin2021}. (c) High-resolution normalized TR-STXM images and line profiles of propagating short-wavelength spin waves in the YIG/Co bilayer at 2.36 GHz with a 10 mT bias field. Figure taken from ref. \cite{Talapatra2023}. (d) Ferromagnetic resonance (FMR) spectra for parallel (black) and antiparallel (red) macrospin states, showing a 1 GHz frequency shift between the states at $H_{\text{ext}} = 0$. An SEM schematic defines the DC external field $H_{\text{ext}}$ and RF field $H_{\text{RF}}$ orientations for the FMR data panels in this figure. Figure taken from ref. \cite{dion2024ultrastrong}. (e) Schematic of the proposed all-on-chip device inspired by previous studies on dynamic dipolar coupling. The Ni\textsubscript{81}Fe\textsubscript{19} (blue), Penrose P3 nanomagnet arrays (maroon), and Pt (gray) are patterned on plain YIG (dark green). The entire chip is placed on an RF co-planar waveguide (CPW). A global spin-pumping voltage can be determined by connecting a nanovoltmeter to the Pt stripes, whereas the local mode profile can be obtained using BLS methods (bright green indicating the laser). 

\textbf{Figure 9.} Future perspective of magnon-magnon coupling as an approach to modulate interactions in other hybrid magnonic systems. Examples including (top to bottom): (a) tuning magnon-photon interaction via magnon-magnon gap, figure taken from Ref. \cite{LiPRR23}; (b) modulating magnon-light interaction and spectrum using magneto-optical effect and its inverse, figure taken from Ref. \cite{inman2022hybrid}; (c) synergy with magnon-exciton modulation, figure taken from Ref. \cite{bae2022exciton}; (d) nonreciprocal magnon-phonon interaction due to magnon-magnon coupling, figure taken from Ref. \cite{MPageSciAdv2020}; (e) artificial spin ice system leveraging magnon-magnon coupling, figure taken from Ref. \cite{dion2024ultrastrong}.

\end{document}